\documentclass{aa}  
\usepackage{graphicx}
\usepackage{txfonts}
\usepackage[colorlinks,linkcolor=blue,citecolor=blue]{hyperref}

\let\vec\undefined
\DeclareMathAccent{\vec}{\mathord}{letters}{"7E}
\usepackage{orcidlink}
\usepackage{natbib}
\usepackage{mathrsfs}
\usepackage[cal=cm]{mathalpha}
\bibpunct{(}{)}{;}{a}{}{,} 
\usepackage[super]{nth}
\usepackage{physics}
\usepackage{ulem}

\makeatletter 
\renewcommand*\sum{\DOTSB\sum@\if@display\limits\fi}
\makeatother  

\begin{document}

	\title{Accretion variability as a probe\\ of black holes and their environments}
	\subtitle{Light curve signatures of a single hot spot}
	\titlerunning{Accretion variability probes of black holes and their environments}
	
	\author{I. Urso
		\inst{1,2}\orcidlink{0009-0003-2105-522X}
		\and C. Deffayet\inst{2}\orcidlink{0000-0002-1907-5606}
		\and A. Held\inst{2,3}\orcidlink{0000-0003-2701-9361} 
		\and
		F. H. Vincent\inst{1}\orcidlink{0000-0002-9214-0830}}
	
	\institute{LIRA, Observatoire de Paris, Université PSL, CNRS, Sorbonne Université, Université Paris Cité, CY Cergy Paris Université, 5 Place Jules Janssen, 92190 Meudon, France\\
		\email{irene.urso@obspm.fr}
		\and
		Laboratoire de Physique de l’ École normale supérieure, ENS, Université PSL,
		CNRS, Sorbonne Université, Université Paris Cité, F-75005 Paris, France 
		\and
		LUX, Observatoire de Paris, Université PSL, CNRS, Sorbonne Université, 5 Place Jules Janssen, 92190 Meudon, France
	}
	
	\date{Received September 15, 1996; accepted March 16, 1997}

	\abstract
	{The compact radio source at the centre of the Milky Way, Sagittarius A*, exhibits temporal broadband brightness variability. The observed timescales of such fluctuations, ranging from minutes to hours on horizon scales, arise from the complex interplay of gravitation, plasma, and light propagation processes in the immediate vicinity of the supermassive black hole. Kinetic and fluid simulations associate this variability with turbulence and the formation of magnetised, localised plasma structures.}
	{Focusing on the case of a single fluctuating compact emission region, a `hot spot', in circular rotation, we study how intrinsic source fluctuations map onto different regions of the observer’s screen. Our aim is to identify distinct imprints of light propagation effects, which encode information about the source’s orientation and velocity, as well as the underlying space-time geometry.}
	{We simulate image time series by propagating the emitted light to a distant observer's sky plane in Minkowski, Schwarzschild, and Kerr space-times. We then compute the corresponding observed light curves and power spectra to characterise their periodic content, both for the full observed images and for selected angular sectors of the observer's sky plane.}
	{We derive a semi-analytical model for the power spectrum of a fluctuating hot spot in flat space-time, reproducing all the qualitative features of curved-space-time spectra at moderate inclinations and providing a physical interpretation of their structure. Moreover, we describe how such power spectra are affected by space-time geometry, inclination, and source velocity, and show that orthogonal two-sector angular decompositions of the sky plane provide information on system orientation and lensing effects.}
	{The physical effects highlighted in this paper are expected to be present in the timing signature of accretion-flow variability, as observed by current instruments as well as future photometric and interferometric facilities.
    Additional work is needed to investigate if our conclusions hold for realistic disc configurations.}
	
	
	\maketitle

	\section{Introduction}
	\begin{figure*}[!t]
		\centering
		\resizebox{14.5cm}{!}
		{\hspace{-0.3cm}
			\includegraphics[width=7.cm]{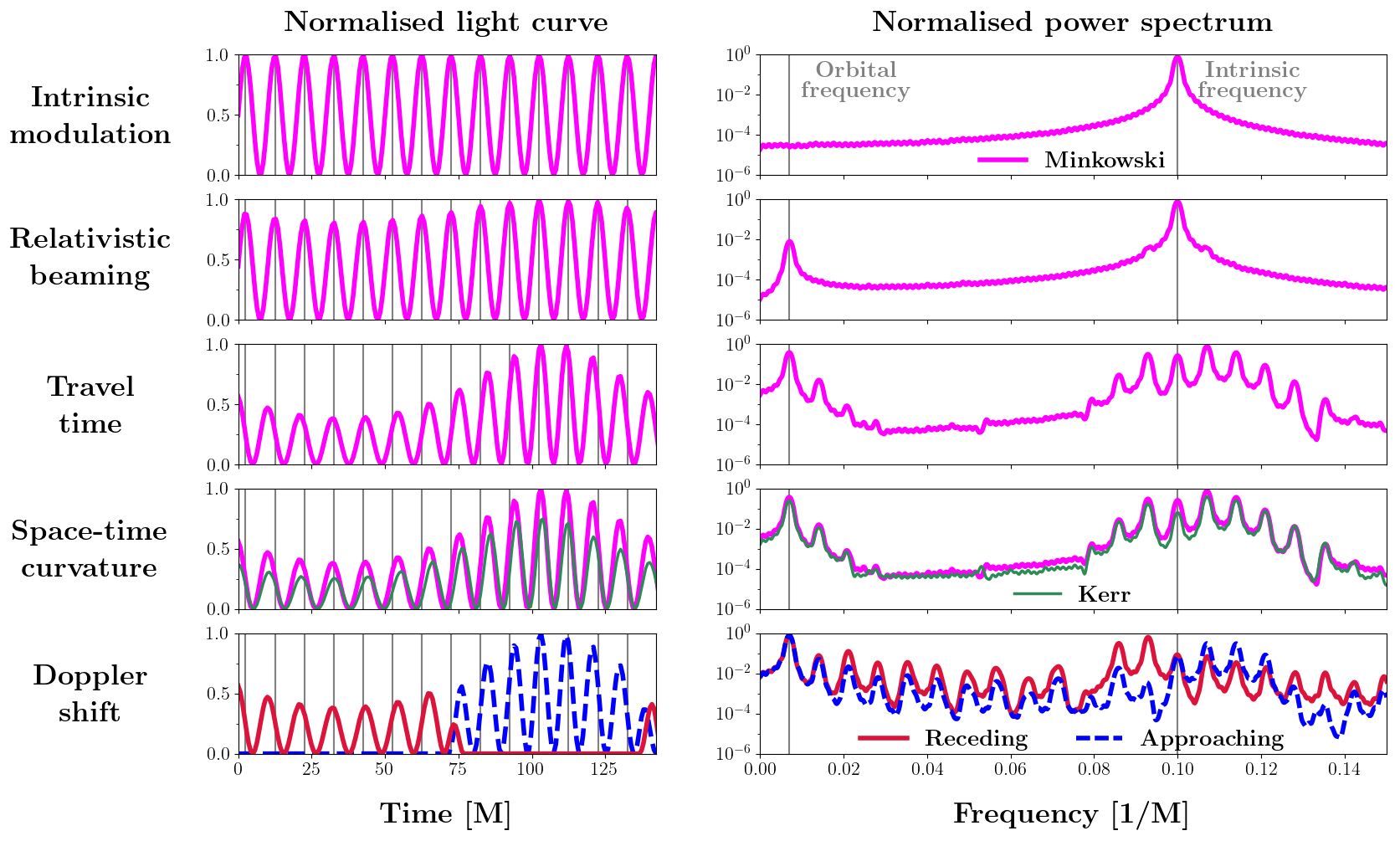}}
		\caption{Light curves (left column) and corresponding Gaussian-filtered power spectral densities (right column) of a sinusoidally emitting source in circular motion, observed at an inclination of $i=160^\circ$, as the ray-tracing treatment and subsequent analysis are progressively refined. Both quantities are normalised to their respective maxima unless stated otherwise. The first two rows assume the fast-light approximation, while the finite speed of light propagation is accounted for from the third row onward. In the first row, the signal contains only the intrinsic source modulation sinusoidal frequency and vertical lines in the first column indicate the maxima of this sinusoid. In the second row, the additional source orbital frequency component is introduced through relativistic beaming. In the third row, accounting for the difference between emission and observation times produces an accordion-like structure in the light curve, which translates into a complex spectral distribution around the two frequencies identified above and marked by vertical lines in the second column. This also leads to changes in the projected area of the source, which modulate the light-curve amplitude and redistribute power across the spectrum. The fourth row overlays the flat space-time signals (thick magenta) with those obtained around a rotating black hole (thin green), all normalised to the maximum values in flat space-time. The frequency content is very similar in the two cases, with the main difference being an overall flux rescaling. Finally, in the last row, the stretched and compressed components of the accordion-like light curve in the complete Minkowski model, collected from regions of the observer's sky plane where the source is receding or approaching, respectively, are isolated in solid red and dashed blue. They yield redshifted and blueshifted spectra, which are normalised to their respective maxima. This paper provides a semi-analytical description of the spectra in flat space-time and shows how information about the system can be extracted from both full light-curve spectra and relative measurements of partial spectra.} 
		\label{fig:Explanatory}
	\end{figure*}
	The compact radio source at the centre of our Galaxy, Sagittarius A* \citep[Sgr A*,][]{Balick1974SgrA}, is widely accepted to be a supermassive black hole (SMBH)
	of mass $M\approx4\times10^6$ $\mathrm{M}_{\odot}$ \citep{Do2019S2} surrounded by an advection dominated, radiatively inefficient accretion flow \citep[ADAF/RIAF,][]{Narayan1995SgrA,Yuan2014ARIAF,EHTVSgrAModels}. At a distance of only $D \approx 8$ kpc \citep{GRAVITY2019Distance}, Sgr A* has the largest apparent angular size among known black holes, and has therefore been the focus of numerous observational campaigns.\\
	\indent Over the past few decades, observations of Sgr A* have revealed significant variability of its luminosity across the entire electromagnetic spectrum, ranging from radio \citep{Brown1982Radio} and X-ray \citep{Baganoff2001Xray}, to near-infrared \citep{Ghenzel2003NIR,Ghez2004IR} and millimetre wavelengths \citep{EHTIVSgrAVariability}. The physical origin of these brightness fluctuations remains debated. Fluctuations in stellar wind feeding from
	the Wolf-Rayet stars plausibly introduce long-timescale variations \citep{Gezel2010SgrAOverview,Ressler2020StellarFeed},
	while turbulence, shocks, or magnetic reconnection triggered at smaller radii in the accretion disc, are likely the causes of shorter-timescale
	fluctuations \citep{Chan2009MRI,Dexter2013Shocks,Ripperda2020HotSpot}. This variability manifests as a continuously and stochastically varying `quiescent
	state', punctuated by prominent brightness enhancements, commonly referred to as `flares' \citep{GRAVITY2020SgrAFlux}.
	\\
	\indent Millimetre synchrotron emission dominates Sgr A* spectral energy distribution \citep{Yuan2003RIAF} and originates predominantly near the black hole event horizon \citep{EHTISgrAShadow}, where hot thermal electrons of the accreting plasma move at relativistic speeds. Timescales associated with horizon-scale processes are then set by the gravitational timescale of Sgr A*, $t_g=GM/c^3\approx20$ s, implying accretion-flow brightness variability of the order of just a few minutes.
	Millimetre observations of Sgr A* total-flux variability confirm
	this expectation, finding a red-noise spectrum across lags from a single
	minute to several hours, with a break to white noise for longer delays \citep{Wielgus2022Millimeter}. Similar behaviour is also detected at near-infrared wavelengths \citep{Do2009NIR}.\\
	\indent Stochastic processes are qualitatively well recovered in state-of-the-art numerical simulations of black hole accretion flows. General relativistic magnetohydrodynamic (GRMHD) models typically produce a red-noise spectrum\footnote{GRMHD models are characterised by a power-law slope of -2 to -3 in frequency and a steepening at high frequencies, in contrast to observations, which show shallower slopes and little evidence of damping.} at millimetre wavelengths \citep{Porth2019GRMHD}, but generally fail to capture secular evolution and its associated low-frequency power. These fluid simulations commonly generate localised and magnetised inhomogeneities, referred to as `hot spots' or `plasmoids' \citep{Nathanail2020Plasmoid,Ripperda2020HotSpot}, whose macroscopic combination is thought to drive flares \citep{Ripperda2022Flare}. Our understanding of their origin and impact is steadily advancing \citep{Ficarra2026Inhomogeneities}. Complementary general relativistic particle-in-cell (GRPIC) models \citep{Mehlhaff2025PIC}, also find power-law power spectra of bolometric luminosity, over frequency ranges overlapping those probed by GRMHD simulations. This broad-band high-frequency variability is associated with magnetic reconnection and the self-similar plasmoid hierarchy it produces.\\
	\indent A growing body of literature investigates how accretion-flow variability can be used to constrain the properties of observed black holes. So far, these studies have focused on probing space-time curvature through autocorrelation functions of intensity fluctuations \citep{Wong2021Glimmer,Bezdvekovz2026Autocorrelations} caused by `secondary' radiation, which loops around the black hole before reaching the observer and gives rise to delayed light echoes of the source \citep{Gralla2020Lensing,Hadar2021PhotonRingAutocorrelation}. However, direct detection of such secondary `glimmers' in
    the total-image light curve at millimetre wavelengths is difficult because the power in echoes is exponentially suppressed relative to the direct emission \citep{Chesler2021Autocorrelations}, and autocorrelation peaks are absent when the intrinsic correlation timescale of the source is greater than the characteristic time lapse between lensed images \citep{Cardenas2024Autocorrelations}. Overcoming these challenges may require visibility-space \citep{Wong2024EchoesVLBI} or  spectrotemporal correlation observables \citep{Hadar2023Spectrotemporal,Harikesh2025Spectrotemporal}, which require, respectively, space-based very long baseline
	interferometers \citep[BHEX,][]{Johnson2024BHEX} or future X-ray spectroscopic missions \citep[NewAthena,][]{Nandra2013NewAthena}.\\
	\indent In this paper, we pursue a complementary and largely unexplored route, investigating light-propagation imprints through fluctuations in the `direct', non-lensed image of a variable emitter. These imprints are summarised in Figure~\hyperref[fig:Explanatory]{\ref{fig:Explanatory}} and detailed below. By complementing the standard analysis of integrated light curves with information on the spatial evolution of the emitting region, our approach is relevant to existing GRAVITY and ALMA observations of flaring activity near the last stable circular orbit of Sgr A*, which combine time-resolved flux measurements with astrometric tracking \citep{GRAVITY2018Flares,Wielgus2022Flare}, as well as to soon-to-be-available horizon-scale movies of accretion-flow dynamics around SMBHs enabled by extensions of the Event Horizon Telescope (EHT) \citep{Doeleman2019MoviesGround,Johnson2019MoviesSpace}. Besides relying only on the routinely detected direct emission, our method is robust in two further respects. First, it does not intrinsically require exceptional flaring events, but can instead target the statistical information contained in the ever-present quiescent emission.  Second, it can be applied to shorter time series than those typically used in autocorrelation studies and can also exploit relative measurements, both of which reduce the reliance on long-term stationarity assumptions. While this work restricts its focus to a single hot spot, identified as a key component of variability in plasma simulations, its results provide a basis for extension to full-disc configurations, which will be explored in subsequent works.\\
	\indent The remainder of this paper is organised as follows. First, Section \hyperref[sec:modelling]{\ref{sec:modelling}} describes the setup and modelling assumptions of our simulations. Then, Section \hyperref[sec:propagation]{\ref{sec:propagation}} details the light-propagation effects that shape the observed variability and constitute the key physical processes investigated in this work. These effects are derived semi-analytically in Minkowski space-time, where the calculations are tractable and capture the dominant features also present in curved space-time (see Figure \hyperref[fig:Explanatory]{\ref{fig:Explanatory}}). Next, in Section \hyperref[sec:4]{\ref{sec:4}}, we show how light curves extracted from different sectors of a time series of images can be used to constrain either the dynamics and orientation of the source at low inclination, or space-time properties at high inclination. Finally, in Section \hyperref[sec:end]{\ref{sec:end}}, we discuss the implications of our findings and provide concluding remarks.

	\section{Modelling}
	\label{sec:modelling}
	
	In this section, we present our model for the equatorial hot spot possessing intrinsic time variability and rotating around the compact object. In addition to presenting the simulation setup, the background space-time geometries, and the physical characteristics of the emitting source, we also comment on the choice of the reference frame used to model astrophysical fluctuations.\\
	\indent Throughout the paper, we employ spherical coordinates $(t,r,\theta,\varphi)$ and geometrized units ($G=c=1$).\footnote{Here, $G$ is the gravitational constant and $c$ the speed of light.} Both times and lengths are then expressed in terms of the black hole mass, $M$.

	\subsection{Ray-traced movies and light curves}

We used the backward ray-tracing code \texttt{GYOTO} \citep[General RelativitY Orbit Tracer of the Observatoire de Paris,][]{vincent2011gyoto} to compute time series of images of a compact hot spot orbiting in the equatorial plane, interpreting each image as the frame of a movie. The movies span $n_\mathrm{orb}=5$ orbits of an emitter with orbital period $T_\mathrm{orb}$, so that their total duration is $\mathcal{T}\equiv n_\mathrm{orb}T_\mathrm{orb}$, and each movie consists of $n\equiv200\,n_\mathrm{orb}$ uniformly distributed frames.\footnote{For a hotspot at radius $r_\mathrm{orb}=8$ $M$, $\mathcal{T}\approx142$ $M$, corresponding to a cadence of $\delta t\approx0.14$ $M$, or about 3 s for Sgr A*.} In this preliminary study, we did not attempt to reproduce realistic observational samplings, but instead opted for a sufficiently long integration time and temporal resolution to ensure robust estimates of the power spectra.\\
\indent Null geodesics were shot from the distant observer’s 2D screen towards the emitter, and the integration was stopped after the first equatorial crossing, so that lensed echoes of the emission are neglected. A point in the light curve is defined as the measured flux, that is, the sum of the intensity of all the pixels in a frame multiplied by the solid angle covered by a pixel, as defined below. The emissivity of the astrophysical emitter orbiting around the central compact object follows a semi-analytical prescription within \texttt{GYOTO}, as described in Section \hyperref[sec:blob]{\ref{sec:blob}}, and we used the `slow-light' paradigm, accounting for
	the finite velocity of light, so that different pixels of the screen typically receive light emitted at different epochs.\\
	\indent The central compact object shares the properties of Sgr A*: it has mass $M=4.3\times10^6$ $\mathrm{M}_{\odot}$ \citep{GRAVITY2022Mass} and it is located at a distance $D=8.28$ kpc from the observer \citep{GRAVITY2021Distance}. We included the inclination, or polar angle, $i=160^\circ$, motivated by GRAVITY astrometric measurements of flares \citep{GRAVITY2018Inclination}. We also define its complementary angle $\iota\equiv\pi-i$, such that the spatial Cartesian coordinates of the observer can be simply expressed as $(x_\mathrm{O},y_\mathrm{O},z_\mathrm{O})\equiv(0,-D\sin\iota,-D\cos\iota)$.\\
	\indent We computed our 2D images, covering a field of view of $\mathcal{F} = 40$ $M$, on a discretised polar grid with $n_{\psi}=360$ equally spaced polar angular directions and $n_{\rho}=512$ uniformly distributed radial points, yielding resolutions $\mathrm{d}\psi=1^\circ$ and $\mathrm{d}\rho\approx0.08$ $M$, respectively. The solid angle subtended by a pixel of coordinates $(\rho,\psi)$ is then:\footnote{Given the extremely small field of view, the angle relative to the screen normal can be neglected.} $\mathrm{d}\Omega=(\rho\,\mathrm{d}\psi\,\mathrm{d}\rho)/D^2$. The parameters of the ray-tracing setup are summarised in Table \hyperref[tab:simu]{\ref{tab:simu}}.
	
	\begin{table}[!ht]
		\caption{Parameters of the ray-tracing setup.}            
		\label{tab:simu}      
		\centering                    
		\begin{tabular}{c c c}        
			\hline\hline\rule{0pt}{2.3ex}
			Symbol & Value & Property\\    
			\hline\rule{0pt}{2.3ex}         
			$D$ & $8.28$ kpc & Black hole distance \\          
			$M$ & $4.3\times10^6$ $\mathrm{M}_{\odot}$& Black hole mass \\  
			$a$ & $\{0;\:0.94\}$ & Black hole spin parameter \\
			$i$ &  $\{160^\circ;100^\circ\}$ & Inclination angle\\ 
			$\mathcal{F}$ &  $40$ $M$ & Field of view \\
			$n_{\psi}$ & 360 & Number of polar points \\
			$n_{\rho}$ & 512 & Number of radial points \\
			$n_\mathrm{orb}$& $5$ & Number of orbits\\
			$n$& $200\,n_\mathrm{orb}$ & Number of movie frames\\
			\hline                        
		\end{tabular}
	\end{table}
	
	\subsection{Space-time geometry}

The investigated light-curve signatures in the direct image are affected not just by the source properties, but also by the space-time geometry. By gradually increasing the geometric complexity, we aimed to isolate and characterise distinct imprints. To this end, we proceeded successively from Minkowski to Schwarzschild and Kerr space-time, using coordinates $(t,r,\theta,\varphi)$, interpreted as spherical, Schwarzschild-Droste, and Boyer-Lindquist coordinates, respectively.\\
\indent We define the dimensionless spin
	parameter as $a\equiv J/M$, where $J$ is the angular
	momentum and $M$ the mass of the black hole.
	The Kerr metric, in Boyer-Lindquist coordinates, reads
	\begin{align}
		\mathrm{d} s^2=&-\left(1-\frac{2Mr}{\rho^2}\right)\,\mathrm{d} t^2-\frac{4aMr\sin^2\theta}{\rho^2}\,\mathrm{d} t\,\mathrm{d}\varphi+\frac{\rho^2}{\Delta}\,\mathrm{d} r^2 \notag\\
		&+\rho^2\,\mathrm{d} \theta^2+\left(r^2+a^2+\frac{2a^2Mr\sin^2\theta}{\rho^2}\right)\sin^2\theta\,\mathrm{d} \varphi^2\:,
	\end{align}
	with 
	\begin{align}
		\rho^2\equiv r^2+a^2\cos^2\theta\quad\text{and}\quad\Delta\equiv r^2-2Mr+a^2\:.
	\end{align}
	The Schwarzschild non-rotating space-time metric, expressed in Schwarzschild-Droste coordinates, is retrieved in the limit $a\to0$. Minkowski space-time, in usual spherical coordinates, is recovered by subsequently taking $M\to0$. For a rotating black hole, we aligned the spin axis with the angular momentum vector of the emitting source and we adopted the high spin value $a=0.94$, following a representative model satisfying EHT observational constraints under the assumption of an external Faraday screen \citep{EHTVIII2024Polarised}.
	
	\subsection{Emitting source}
	
	In this paper, we model the emitting region as a compact hot spot orbiting the black hole on a circular equatorial trajectory. This simplified setup provides a controlled framework for a partially analytical treatment of relativistic propagation effects.
	
	\subsubsection{Circular motion}
	The emitter in our simulations moves along circular orbits in the equatorial plane and its observed motion is clockwise.
	In the flat space-time case, the orbital dynamics is governed by a Newtonian central point mass, $M$, which does not curve space-time. If a black hole with positive spin is present, our choices imply that the orbit of the emitting source is prograde.\\
	\indent We introduce the angular velocity $\Omega_\mathrm{orb}\equiv u^\varphi/u^t$, related to the orbital period, $T_\mathrm{orb}$, and the orbital frequency, $f_\mathrm{orb}$, through $\Omega_\mathrm{orb}=2\pi/T_\mathrm{orb}=2\pi f_\mathrm{orb}$. The 4-velocity describing a circular orbit is completely\footnote{The time component $u^t$ is found via the normalisation $u^\mu u_\mu=-1$.} determined by $\Omega_\mathrm{orb}$ and it reads
	\begin{align}
		u^\mu&=u^t(1,0,0,\Omega_\mathrm{orb})\:.
	\end{align}
 We chose $\Omega_\mathrm{orb}$ to match the Newtonian result for a classical two-body problem, rather than the specific circular-geodesic expressions of each space-time. It is given by\footnote{This coincides with the Keplerian Schwarzschild angular velocity  \citep{Bardeen1972Kerr}.}
 \begin{align}
 	\Omega_\mathrm{orb}=\frac{\sqrt{M}}{r^{3/2}}\:.\label{eq:Omega}
 \end{align}
 This simple choice was made both to avoid model-dependent spin impacts that could lead to misleading conclusions on the variability observables, and by the absence of a consensus prescription for the velocity law governing hot-spot motion \citep{Aimar2023Flares}. Finally, we only considered sources orbiting outside of the Schwarzschild innermost stable circular orbit (ISCO), that is, $r>6M$, which trivially guarantees subluminal motion in both the Minkowski and Schwarzschild cases, and we have numerically verified that the 4-velocity remains strictly time-like, $u^\mu u_\mu < 0$, across the considered Kerr trajectories despite the non-geodesic nature of the chosen velocity profile.
	
\subsubsection{Hot spot}
\label{sec:blob}
	
	At every point along its orbit, the hot spot was defined as a disc in the equatorial plane, centred at spatial Cartesian coordinates $(x_\mathrm{c},y_\mathrm{c},z_\mathrm{c}) =
	(r_\mathrm{c} \cos\varphi_\mathrm{c} ,r_\mathrm{c} \sin\varphi_\mathrm{c},0)$, and with spatial extent set by the Euclidean distance from this centre, $d_\mathrm{spot}=1$ $M$.\\
	\indent The centre was required to follow a circular time-like geodesic at constant radial distance, $r_\mathrm{c}=r_\mathrm{orb}$, meaning that the $\varphi$-coordinate could be expressed as $\varphi_\mathrm{c}=\Omega^\mathrm{c}_\mathrm{orb}t+\varphi^{\,c}_0$, where $\Omega^\mathrm{c}_\mathrm{orb}\equiv\sqrt{M}/r_\mathrm{c}^{3/2}$ and $\varphi^{\,c}_0$ denotes the initial azimuthal position of the centre.
		The disc-like hot spot defined above undergoes solid-body circular rotation. We adopted $r_\mathrm{orb}=8$ $M$, a value close to the orbital radius inferred for Sgr A* flares \citep{GRAVITY2018Flares,Wielgus2022Flare} and which ensures all constituent points of the hot spot lie at radii larger than $r_\mathrm{ISCO}$. The initial 4-position of the hot spot was chosen such that it appears on sky along the projected angular-momentum vector of the source in the first time frame.\\ 
	\indent Time modulations were added by multiplying the original unitary emissivity of the hot spot by a sinusoidal factor with positive offset and frequency $f_\mathrm{mod}$. For a perfectly equatorial source, the emissivity coincides with the emitted intensity, denoted $I^\mathrm{em}$, which is then given by
	\begin{align}
		I^\mathrm{\,em}=1+\sin(2\pi\,f_\mathrm{mod}\,t^\mathrm{\,em})\:.
    \label{eq:emissivity}
	\end{align}
	We stress that the modulation was defined with respect to the global coordinate time, as justified in the next section, and evaluated at the coordinate emission time. The emitted intensity of the hot spot appears then to vary between zero and two, and the corresponding modulation frequency, $f_\mathrm{mod}=0.1$ $M^{-1}$, was selected within the range of frequencies found in both GRMHD and GRPIC simulations \citep{Porth2019GRMHD,Mehlhaff2025PIC}. The numerical values of the hot-spot model are summarised in Table \hyperref[tab:blob]{\ref{tab:blob}}.
	\indent  
		\begin{table}[!ht]
		\caption{Parameters of the hot spot.}            
		\label{tab:blob}      
		\centering                    
		\begin{tabular}{c c c}        
			\hline\hline\rule{0pt}{2.3ex}     
			Symbol & Value & Property\\    
			\hline\rule{0pt}{2.3ex} 
			$d_\mathrm{spot}$ & $1$ $M$ & Hot spot extension \\
			$r_\mathrm{orb}$ & $8$ $M$ & Hot spot orbital radius \\      
			$f_\mathrm{mod}$ & $0.1$ $M^{-1}$ & Hot spot emissivity frequency\\
			$\varphi^\mathrm{c}_0$ & $-90^\circ$ & Hot spot initial azimuthal position\\[2pt]
			\hline                
		\end{tabular}
	\end{table}
	
	\subsection{A note on time modulations}
	\label{sec:modulation}
	
	In equation \eqref{eq:emissivity}, we defined the time modulation with respect to the global coordinate time, $t$,\footnote{Depending on the space-time considered, $t$ denotes either the Minkowski time coordinate or the Boyer-Lindquist time coordinate.} rather than the proper time, $\tau$, along the hot spot’s worldline. In this simplified configuration, kinematic time-dilation effects are ignored. However, Rømer (phase) delay, lensing effects, and classical Doppler (frequency) shifts are fully incorporated through the transformation between emission and arrival times, which accounts for both the finiteness of the speed of light, $c$, and the relative motion between source and observer, that is, variations in their separation.\\
	\indent In a special relativistic framework, the Lorentz factor of the rigidly rotating hot spot is $\gamma=1/\sqrt{1-(v_\mathrm{c}/c)^2}>0$, where $v_\mathrm{c}\equiv r_\mathrm{c}\,\Omega_\mathrm{orb}^\mathrm{c}$ is the orbital velocity measured by a static observer at the location of the hot spot.\footnote{In flat space-time, such an observer is inertial and shares the same global time coordinate as a distant static observer, so this Lorentz factor directly relates the hot spot’s proper time to the distant observer’s time.} The proper time interval measured along the worldline of the hot spot, $\differential \tau$, and the proper time interval measured by a distant static observer, $\differential t$, are related by $\differential \tau = \differential t/\gamma$. Since $\gamma$ is constant for motion at constant velocity, this integrates to $\tau=t/\gamma$.\footnote{Similarly one has $\phi=\gamma\varphi$, with $\phi$ the proper azimuthal coordinate of the hot spot. This implies that, in our model, the hot spot is not circular in its rest frame.} Imposing our modulation in coordinate time therefore amounts to prescribing a frequency $f_\mathrm{mod}=f_\mathrm{proper}/\gamma$, where $f_\mathrm{proper}$ is the intrinsic modulation frequency in the hot spot’s rest frame. In other words, the frequency $f_\mathrm{mod}$ corresponds to the time-dilated value of the proper frequency. When light-travel-time effects are additionally taken into account, the observed frequency further includes the usual Doppler factor. In terms of $f_\mathrm{proper}$, this yields the full relativistic Doppler shift. Note, by the way, that our choice of value for $f_\mathrm{mod}$ is justified because GRMHD and GRPIC simulations characterise temporal variability in terms of coordinate time.\footnote{Typically, Kerr-Schild coordinates are used, which reduce to Boyer-Lindquist coordinates for circular motion at constant radius.}\\
	\indent In summary, using the coordinate time $t\,$ to define the modulation is conceptually sound within our setup. However, one must remain aware of the reference quantity when interpreting the observed frequency, especially when attempting to infer physical properties of the emitting source. 
    Besides, the simplification adopted here no longer holds when $\gamma$ is not constant, as occurs, for instance, when a radial velocity component is added to the circular motion. In that case, the comparison between the observed frequency and either $f_\mathrm{mod}$ or $f_\mathrm{proper}$ would lead to qualitatively different behaviours, rather than a simple rescaling.

	\section{Propagation effects in flat space-time}
	\label{sec:propagation}
\begin{figure}[ht!]
	\centering
	\includegraphics[width=8.5cm]{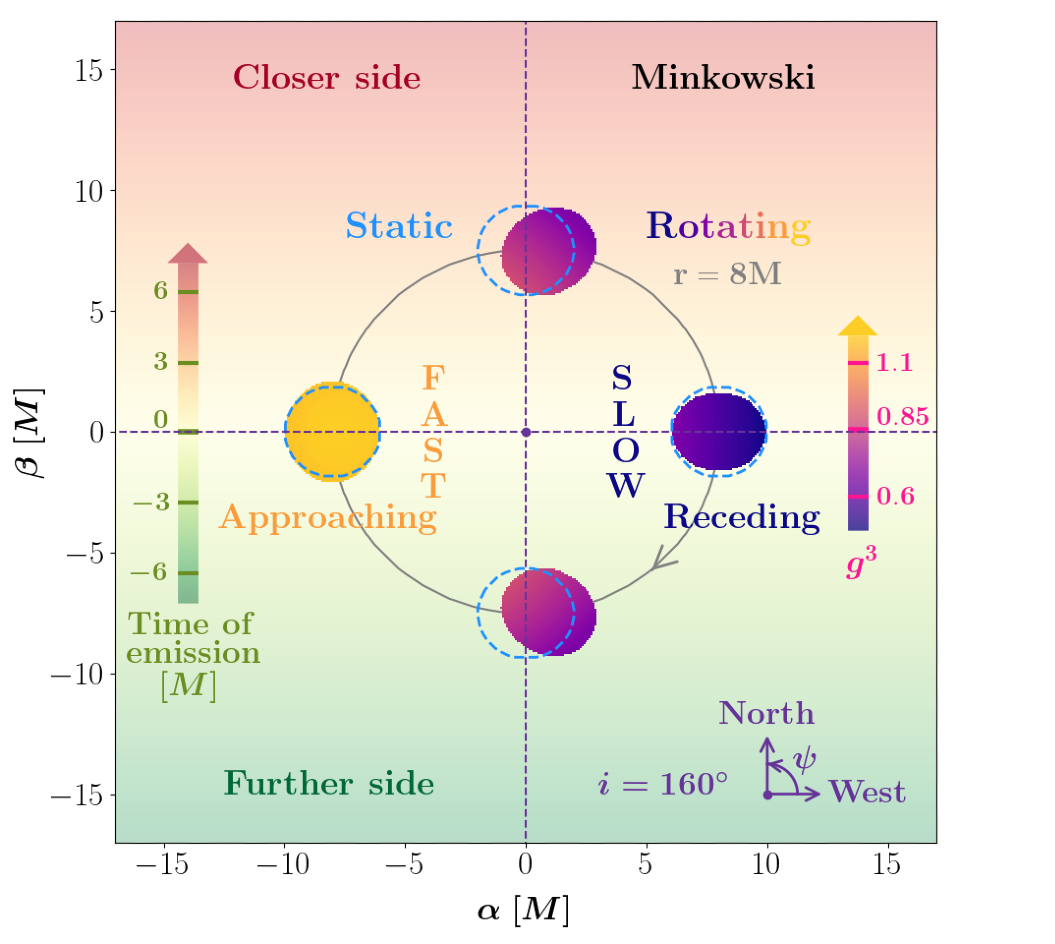}
	\caption{Light propagation effects on the observed hot spot following a circular orbit in Minkowski space-time. Its apparent motion is clockwise and four projections, imaged at fixed observing time delay $\Delta t^\mathrm{obs} = T_\mathrm{orb}/4$, are presented in solid colours. On these four hot-spot snapshots, the right colour scale indicates relativistic beaming, which contributes to the slow modulation observed in the light curves of Figure \hyperref[fig:Explanatory]{\ref{fig:Explanatory}}, while the fast modulation is associated with the intrinsic hot-spot variability, which is not represented here. The observer's inclination, here $i=160^\circ$, causes different emission times to be imaged simultaneously, as indicated by the background colour map with colour scale on the left. Inclination-induced projection contractions already deform the image of circular static emitters (dashed light-blue lines), with doubled $d_\mathrm{spot}$ for visualisation purposes. The hot spot's motion introduces additional distortions that combine with relativistic beaming, providing a second contribution to the slow modulation of the light curves. The combined effects of observer inclination, source velocity, and the finite speed of light additionally lead to an apparent non-constant angular velocity of the imaged hot spot, which translates into the accordion-like behaviour of the light curves.} 
	\label{fig:starting_point}
\end{figure}

Here, we detail light propagation and its effect on observed modulated light curves in Minkowski space-time, where analytical derivations provide a convenient framework for interpreting the outputs of numerical simulations. This choice is primarily motivated by the strong similarities between light curves in flat and curved space-times at low to moderate inclinations, as shown in the fourth row of Figure \hyperref[fig:Explanatory]{\ref{fig:Explanatory}}. Throughout this section, numerical results in Minkowski, Schwarzschild, and Kerr space-times are superposed for comparison.\footnote{The analytical predictions refer to the hot-spot centre, whereas the numerical comparison uses the barycentre of the imaged blob in the Cartesian screen coordinates $(\alpha,\beta)$, shown in Figure \hyperref[fig:starting_point]{\ref{fig:starting_point}}. Although it does not necessarily coincide with the hot-spot centre, this barycentre provides a reliable proxy for its position, as confirmed by the agreement with the analytical expectations.} The aspects modified by space-time curvature are discussed in the next section.\\
\indent The observed imprints of light propagation arise from the mapping between the observation coordinates on the screen and the impact coordinates of the corresponding light rays at the source. Complete emission maps for the coordinates $(t,r,\varphi)$, assuming a fixed polar angle $\theta=\pi/2$, are provided in Appendix \hyperref[app:hit]{\ref{app:hit}}. A comprehensive illustration of the key light-propagation effects for the hot spot, on a circular orbit observed at $i=160^\circ$ in flat space-time, is presented in Figure \hyperref[fig:starting_point]{\ref{fig:starting_point}}.\\
\indent Here, in order to provide intuitive explanations, we require the vertical axis of the screen to be aligned with the projected angular-momentum vector of the source, but the results presented can be generalised to whatever orientation of the observer's screen. Finally, since we deal with time-dependent signals, for practicality, we introduce the following notation for a generic quantity, $Q$, decomposed as
\begin{align}
	Q=Q_0+Q_t\:,\label{eq:notation}
\end{align} 
into a static component, $Q_0$, and a time-varying component, $Q_t$.

\subsection{Light travel-time delays}
\label{sec:times}
\begin{figure*}[ht!]
	\centering
	\resizebox{15cm}{!}
	{\includegraphics{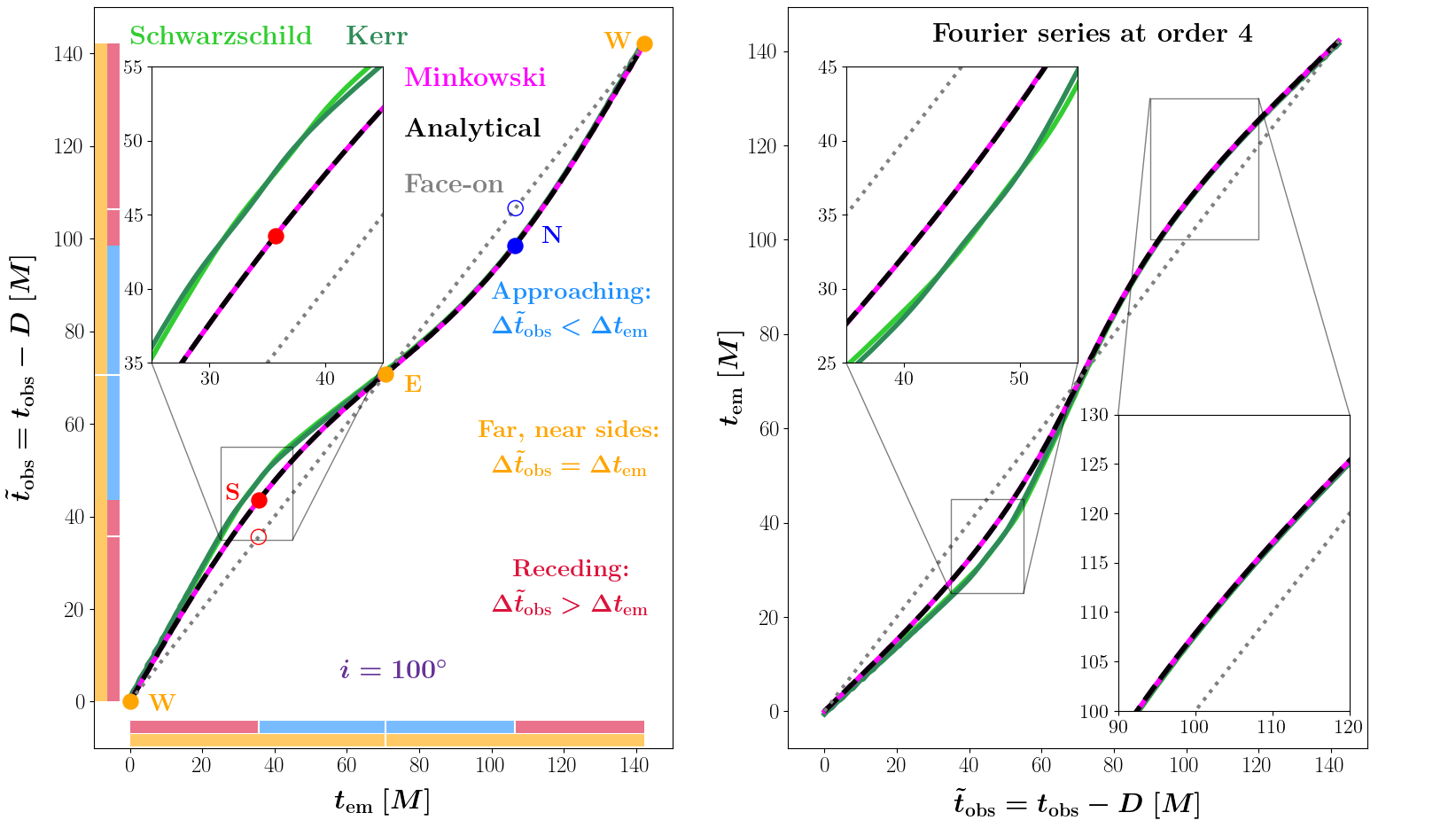}}
	\caption{Relations between observing and emission times: $\tilde{t}_\mathrm{obs}\equiv t_\mathrm{obs}-D$ as a function of $t_\mathrm{em}$ (left panel) and $t_\mathrm{em}$ as a function of $\tilde{t}_\mathrm{obs}$ (right panel). Results are shown at $i=100^\circ$ for Minkowski (magenta), Schwarzschild (light green), and Kerr ($a=0.94$, dark green) space-times. Within flat space-time, the dashed black curves represent the matching analytical predictions \eqref{eq:flight} and \eqref{eq:FourierSeries} and the dotted grey lines correspond to a hot spot observed face-on. Additionally, for the Minkowski case, filled coloured points indicate the hot spot at its cardinal positions on the screen, while empty points denote the corresponding positions on the screen of a face-on observer. The coloured bars illustrate how equal intervals of emission time are observed to be stretched when the source is receding, or compressed when it approaches the observer. In contrast, the time spent by the hot-spot image in the perpendicular halves of the screen is the same in both emission and observation time, and equals $T_\mathrm{orb}/2$.}
	\label{fig:inversion}
\end{figure*}
\indent Because of inclination, in our simulation setup, the coordinate distance to the equatorial orbital plane is smaller for pixels making up the north side of the observer's screen than for those of the south part. In other words, as one image frame is observed at constant observation time, $t^\mathrm{\,obs}$, the radiation reaching the southern and northern sides of the screen must have been emitted earlier and later, respectively, in coordinate time.\footnote{This behaviour can only be captured within the slow-light paradigm.}\\
\indent Mapping the evolution of the emission time of a source with respect to the observer time is essential for interpreting time-dependent phenomena. The relation between the two is given by the time of flight, $F$, which varies from pixel to pixel:
\begin{align}
	t^\mathrm{\,obs}=t^\mathrm{\,em}+F.
	\label{eq:time}
\end{align}
In Minkowski space-time, the time of flight between the observer at $(x_\mathrm{O},y_\mathrm{O},z_\mathrm{O})$ and the hot spot's centre at $(x_\mathrm{c},y_\mathrm{c},z_\mathrm{c})$ is simply given by their Euclidean distance. Thus, defining a rescaled observing time as $\tilde{t}^\mathrm{\,obs} \equiv t^\mathrm{\,obs} - D$ and applying the approximation $D\gg r_\mathrm{c}$, relation \eqref{eq:time} becomes
\begin{align}
	\tilde{t}^\mathrm{\:obs,\,c}_\mathrm{Minko}= t^\mathrm{\,em,\,c}_\mathrm{Minko}+r_\mathrm{c}\sin\iota\sin\,\left(\Omega^\mathrm{c}_\mathrm{orb}t^\mathrm{\,em,\,c}_\mathrm{Minko}+\varphi^\mathrm{\,c}_0\right)
	\label{eq:flight}\:.
\end{align}

The quantity $\tilde{t}^\mathrm{\,obs,\,c}_\mathrm{Minko}$ can be expressed directly in terms of $t^\mathrm{\,em,\,c}_\mathrm{Minko}$. Plotting the former against the latter yields a curve that deviates increasingly from a straight line as the inclination grows, as shown in the left panel of Figure \hyperref[fig:inversion]{\ref{fig:inversion}}. Denoting by $\varphi^\mathrm{\,em,\,c}_\mathrm{Minko}$ the argument of the sine in equation \eqref{eq:flight}, the rescaled time-of-flight term vanishes for $\varphi^\mathrm{\,em,\,c}_\mathrm{Minko}\in\{0;\pi\}$, reaches its maximum at $\varphi^\mathrm{\,em,\,c}_\mathrm{Minko}=\pi/2$, where emitted photons map to the southern axis of the screen, and its minimum $\varphi^\mathrm{\,em,\,c}_\mathrm{Minko}=3\pi/2$, corresponding to the northern screen's axis. 
This transformation implies that uniform intervals in emission time appear stretched or compressed, depending on whether the source is receding from or approaching the observer (see the coloured bars in left panel of Figure \hyperref[fig:inversion]{\ref{fig:inversion}}). In particular, there is a distinction between a horizontal and a vertical split on the screen. If the source remains visible for an equal duration $T_\mathrm{orb}/2$ in the upper and lower halves of the screen, the corresponding interval of observation time is longer on the right-hand receding side and shorter on the left-hand approaching side. Moreover, the observation time accumulated over an orbital interval, $\Delta \tilde{t}^\mathrm{\,obs,\,c}_\mathrm{Minko}=\tilde{t}^\mathrm{\,obs,\,c}_\mathrm{Minko}(\varphi^\mathrm{\,em,\,c}_\mathrm{Minko}=\varphi_f)-\tilde{t}^\mathrm{\,obs,\,c}_\mathrm{Minko}(\varphi^\mathrm{\,em,\,c}_\mathrm{Minko}=\varphi_i)$, is invariant under the transformation $(\varphi_i,\varphi_f)\to(2\pi-\varphi_f,2\pi-\varphi_i)$ because of the sinusoidal dependence of the time-of-flight correction in Minkowski space-time. This implies that the time spent on the left- and right-hand sides is equally distributed between the upper and lower quadrants (the various red and light-blue bar segments in Figure \hyperref[fig:inversion]{\ref{fig:inversion}} share the same length), with the corresponding intervals given by $T_\mathrm{orb}/4\mp  r_\mathrm{c}\sin\iota$, respectively. As a result, the apparent motion is slower on the right and faster on the left, even though the motion in the equatorial plane is characterised by a constant angular velocity.  
These effects are readily visible in Figure \hyperref[fig:starting_point]{\ref{fig:starting_point}}, where four different images of the orbiting hot spot are collected with a fixed delay, $\Delta t^\mathrm{\,obs,\,c}_\mathrm{Minko} = T_\mathrm{orb}/4$.\\
\indent Expressing $t^\mathrm{\,em,\,c}_\mathrm{Minko}$ in terms of $\tilde{t}^\mathrm{\,obs,\,c}_\mathrm{Minko}$ is less straightforward since, for $\iota\neq0^\circ$,\footnote{When $\iota=0^\circ$, $F^\mathrm{\,c}_\mathrm{Minko}= D$ and $t^\mathrm{\,em,\,c}_\mathrm{Minko}=\tilde{t}^\mathrm{\:obs,\,c}_\mathrm{Minko}$.} relation \eqref{eq:flight} is a transcendental equation that cannot be solved algebraically for $t^\mathrm{em,\,c}_\mathrm{Minko}$. However, as we show in Appendix \hyperref[app:Minko]{\ref{app:Minko}}, an exact analytical inversion is provided by a Fourier series:
\begin{subequations}\label{eq:FourierSeries}
	\begin{align}
		&t^\mathrm{\,em,\,c}_\mathrm{Minko}= \tilde{t}^\mathrm{\:obs,\,c}_\mathrm{Minko}+\frac{2}{\Omega^\mathrm{c}_\mathrm{orb}}\sum_{n=1}^\infty\frac{(-1)^n}{n}J_n(nK)\sin\left[\theta_n\left(\tilde{t}^\mathrm{\:obs,\,c}_\mathrm{Minko}\right)\right]\:,\\
		&K\equiv\Omega^\mathrm{c}_\mathrm{orb}\,r_\mathrm{c}\sin\iota\:,\quad\theta_n\equiv2n\pi f^\mathrm{\,c}_\mathrm{orb}\,\tilde{t}^\mathrm{\:obs,\,c}_\mathrm{Minko}+n\varphi^\mathrm{\,c}_0\:,
	\end{align}
\end{subequations}
where $J_n$ is the $n$-th Bessel function of the first kind.
In Figure \hyperref[fig:inversion]{\ref{fig:inversion}}, we superpose solutions \eqref{eq:flight} and \eqref{eq:FourierSeries} on the numerical results.\\
\indent From equations \eqref{eq:emissivity} and \eqref{eq:FourierSeries}, the emitted intensity of the hot spot's centre in Minkowski space-time reads
\begin{subequations}\label{eq:multiPM}
	\begin{align}
		&I_\mathrm{Minko}^\mathrm{\,em,\,c}=1+\sin\left(2\pi f_\mathrm{mod}\tilde{t}^\mathrm{\:obs,\,c}_\mathrm{Minko}+\varphi^\mathrm{\,c}_0+\sum^\infty_{n=1}\Phi_n\sin\left[\theta_n\left(\tilde{t}^\mathrm{\:obs,\,c}_\mathrm{Minko}\right)\right]\right)\:,\\
		&\beta_n\equiv\frac{2}{\Omega^\mathrm{c}_\mathrm{orb}}\frac{(-1)^n}{n}J_n\left(nK\right)\,,\quad\Phi_n\equiv2\pi f_\mathrm{mod}\,\beta_n\:.
	\end{align}
\end{subequations}
The emitted intensity therefore exhibits a harmonic phase-modulation behaviour, where the frequency $f_\mathrm{mod}$ is periodically frequency-shifted by a modulation at the fundamental frequency $f^\mathrm{\,c}_\mathrm{orb}$. The two frequencies satisfy $f_\mathrm{mod}\gg f^\mathrm{\,c}_\mathrm{orb}$.\footnote{In signal-processing terminology, $f_\mathrm{mod}$ and $f^\mathrm{\,c}_\mathrm{orb}$ would be referred to as the carrier and message frequencies, respectively.} As shown in Appendix \hyperref[app:PM]{\ref{app:PM}}, this phase-modulated signal can be rewritten as
\begin{subequations}
	\label{eq:Iem}
	\begin{align}
		&I_\mathrm{Minko}^\mathrm{\,em,\,c}= 1 + \sum_{\boldsymbol{k}\in\mathbb{Z}^\infty} A_{\boldsymbol{k}} \sin(2\pi\left[ f_\mathrm{mod}+m_{\boldsymbol{k}}f_\mathrm{orb}^\mathrm{\,c}\right]\tilde{t}^\mathrm{\:obs,\,c}_\mathrm{Minko}+\Delta\phi)\:,\\
		&A_{\boldsymbol{k}}\equiv\prod_{n=1}^{\infty} J_{k_n}(\Phi_n)\:,\quad m_{\boldsymbol{k}}\equiv \sum_{n=1}^\infty n k_n\:,\quad \Delta \phi\equiv [m_{\boldsymbol{k}}+1]\,\varphi^c_0\:.
	\end{align}
\end{subequations}
where $\boldsymbol{k}$ denotes an infinite integer multi-index, so that $\sum_{\boldsymbol{k}\in\mathbb{Z}^\infty}$ represents the multiple sum $\sum_{k_1, k_2, \dots = -\infty}^{\infty}$. As $f_\mathrm{orb}^\mathrm{\,c}$ is not a multiple of $f_\mathrm{mod}$, every term in the sum carries a non-zero frequency information.\footnote{Given the exponential decay of Bessel functions with index and the relation $f_\mathrm{mod}\gg f^\mathrm{\,c}_\mathrm{orb}$, negative frequency contributions can be safely neglected when computing the power spectral density.} Following the notation \eqref{eq:notation}, the sum appearing in \eqref{eq:Iem} corresponds then to the time-dependent component $I_\mathrm{Minko,\,t}^\mathrm{\,em,\,c}$, while the constant term $I_\mathrm{Minko,\,0}^\mathrm{\,em,\,c}$ is simply equal to 1. With these identifications, we can write
\begin{align}
	I_\mathrm{Minko}^\mathrm{\,em,\,c}= I_\mathrm{Minko,\,0}^\mathrm{\,em,\,c} + I_\mathrm{Minko,\,t}^\mathrm{\,em,\,c}\:.
\end{align}

\subsection{Relativistic beaming}
\label{sec:beam}
\begin{figure}[ht!]
	\centering
	\includegraphics[width=9.5cm]{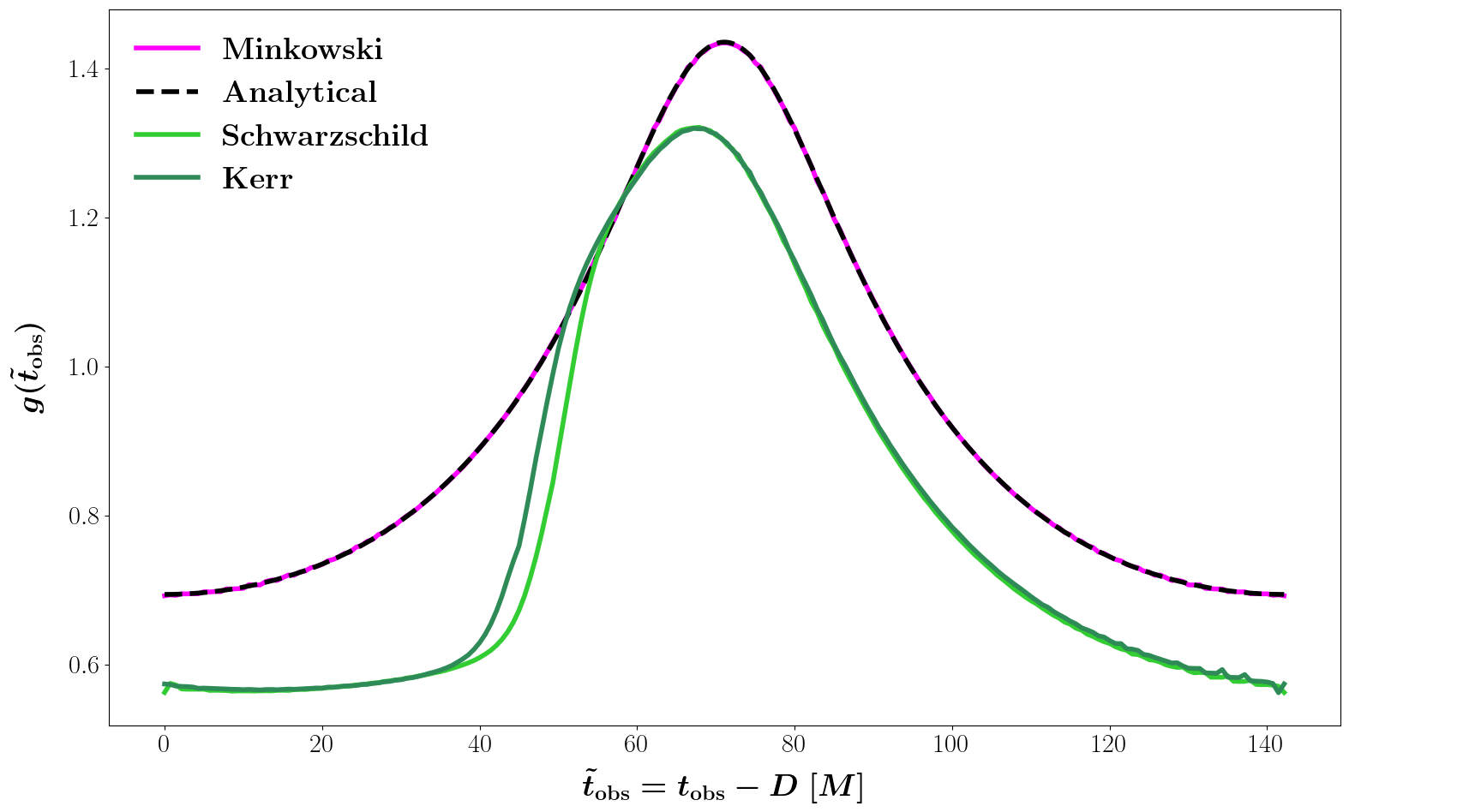}
	\caption{Redshift profile of the hot spot's centre in Minkowski (magenta), Schwarzschild (light green), and Kerr ($a=0.94$, dark green) space-times at $i=100^\circ$.  Within flat space-time, the dashed black curve represents the concordant analytical prediction \eqref{eq:redshiftMinko}.}
	\label{fig:redshift}
\end{figure}

Liouville's theorem states the conservation of phase-space volume: $I_\nu/\nu^3$ is conserved along a geodesic (see equation 22.49 in \citet{MTW1973Gravitation}), with $I_\nu$ the intensity per unit frequency $\nu$.\footnote{By contrast, in Newtonian physics the specific intensity is conserved along rays in free space, $I_\nu=\text{cst}$ (see equation 1.11 in \citet{Rybicki&Lightman1979Radiative}), then no change in apparent brightness occurs.} 
Thus, any specific emitted intensity, $I^\mathrm{em}_\nu$, is related to a respective observed intensity, $I^\mathrm{obs}_\nu$, by \citep{Lindquist1966Transfer}
\begin{align}
	I^\mathrm{obs}_\nu=g^3 I^\mathrm{em}_\nu\:,
	\label{eq:IemIobs}
\end{align}
where $g$ represents the redshift factor:
\begin{align}
	g\equiv\frac{\nu^\mathrm{obs}}{\nu^\mathrm{em}}=\frac{\boldsymbol{p}^\mathrm{obs}\cdot \boldsymbol{u}^\mathrm{obs}}{\boldsymbol{p}^\mathrm{em}\cdot \boldsymbol{u}^\mathrm{em}}\:,
	\label{eq:redshift}
\end{align}
defined as the ratio of observed to emitted frequencies, $\nu^{\mathrm{obs/em}}$, or, equivalently, the ratio of local scalar products between the photon four-momentum, $\boldsymbol{p}^{\mathrm{obs}/\mathrm{em}}$, and the four-velocity of the observer and emitter, $\boldsymbol{u}^{\mathrm{obs}/\mathrm{em}}$, respectively.
Equation \eqref{eq:redshift} leads to the so-called relativistic effect, which modifies the apparent brightness of a moving source: when the emitter moves toward the observer it appears brighter than in the static case, whereas it appears fainter when receding. Beaming effects are shown by the right colour bar in Figure \hyperref[fig:starting_point]{\ref{fig:starting_point}}.\\
\indent The redshift equation \eqref{eq:redshift} of the hot spot's centre, represented in Figure \hyperref[fig:redshift]{\ref{fig:redshift}}, reduces to the following expression in Minkowski space-time, as detailed in Appendix \hyperref[app:Minko]{\ref{app:Minko}}:
\begin{subequations}
\label{eq:redshiftMinko}
\begin{align}
	&g_\mathrm{Minko}^\mathrm{\,c}=\sqrt{\frac{r_\mathrm{c}-M}{r_\mathrm{c}}}\left[1+x \right]^{-1},\\
	&C_2\equiv\sqrt{\frac{M}{r_\mathrm{c}}}\sin\iota\:,\quad x\equiv C_2\cos\left(\Omega_\mathrm{orb}^\mathrm{c}t^\mathrm{\,em,\,c}_\mathrm{Minko}+\varphi_0^\mathrm{\,c}\right).
\end{align}
\end{subequations}
\indent Inserting equation \eqref{eq:FourierSeries}, $x$ can be rewritten as:
\begin{subequations}
	\label{eq:x}
	\begin{align}	
		&x= C_2\cos\left(2\pi f_\mathrm{orb}^\mathrm{\,c}\tilde{t}^\mathrm{\:obs,\,c}_\mathrm{Minko}+\varphi_0^\mathrm{\,c}+\sum^\infty_{n=1}\bar{\Phi}_n\sin\left[\theta_n\left(\tilde{t}^\mathrm{\:obs,\,c}_\mathrm{Minko}\right)\right]\right)\:,\\
		&\bar{\Phi}_n\equiv2\pi f_\mathrm{orb}^\mathrm{\,c}\,\beta_n\:
	\end{align}
\end{subequations}
and expanded similarly as before, as elaborated in Appendix \hyperref[app:PM]{\ref{app:PM}}:
\begin{subequations}
	\label{eq:xPM}
	\begin{align}	
		&x=C_2\sum_{\boldsymbol{k}\in\mathbb{Z}^\infty} \bar{A}_{\boldsymbol{k}}\cos(2\pi[m_{\boldsymbol{k}}+1] f_\mathrm{orb}^\mathrm{\,c}\,\tilde{t}^\mathrm{\:obs,\,c}_\mathrm{Minko}+\Delta\phi)\:,\\
		&\bar{A}_{\boldsymbol{k}}\equiv \prod_{n=1}^{\infty} J_{k_n}(\bar{\Phi}_n)\:.
	\end{align}
\end{subequations}

Beaming effects in Minkowski space-time can then be re-expressed by expanding the cube of \eqref{eq:redshiftMinko}, following equation \eqref{eq:IemIobs}, for small $x$ using the generalised binomial theorem. This yields a non-linear series in powers of a self-phase modulated sinusoids at frequency $f^\mathrm{\,c}_\mathrm{orb}$:
\begin{subequations}
	\begin{align}
		\left(g^\mathrm{\,c}_\mathrm{Minko}\right)^3&=\left[\frac{r_\mathrm{c}-M}{r_\mathrm{c}} \right]^{3/2}\sum_{n=0}^{\infty} \binom{-3}{n}\, x^n\label{eq:beaming}\\
		&=G^\mathrm{\,c}_\mathrm{Minko,\, 0}+G^\mathrm{\,c}_\mathrm{Minko,\,t}\:,
	\end{align}
\end{subequations}
where in the second line, following the notation \eqref{eq:notation}, we have decomposed the expansion into a static component, $G^\mathrm{\,c}_\mathrm{Minko,\,0}$,
and an oscillating component in time, $G^\mathrm{\,c}_\mathrm{Minko,\,t}$. The zero-frequency term includes the $n=0$ mode along with the time-independent contributions appearing when developing higher-order sinusoidal powers using product-to-sum trigonometric identities.

\subsection{Inclination projections and Lampa-Terrell-Penrose effect}
\label{sec:terrell}
The combination of orbital inclination and the relativistic motion of the source leads to significant apparent shape deformations and therefore to flux variations, as the collected flux is proportional to the apparent size of the hot spot on screen.\\ 
\indent Inclination alone induces projection distortions, which cause circular rings in the equatorial plane, such as the orbital trajectory or the edges of a static hot spot, to appear increasingly deformed on the observer’s screen as the inclination approaches edge-on. In Minkowski space-time, they become ellipses with semi-major axis equal to to their radial extension in the equatorial plane and semi-minor axis shrunk by a factor $\cos\iota$\footnote{\label{lab:approx}After applying the small-angle approximation $|\alpha|, |\,\beta| \ll \iota$.} (see dashed contours in Figure \hyperref[fig:starting_point]{\ref{fig:starting_point}}).\\
\indent If the considered hot spot is moving, as for the full-colour rotating hot spot represented in Figure \hyperref[fig:starting_point]{\ref{fig:starting_point}}, its shape on the observer's screen appears further deformed, producing an asymmetry in the observed brightness, even when neglecting beaming effects (see Figure \hyperref[fig:terrell]{\ref{fig:terrell}}). The deformations induced by the motion of the emitter are amplified at higher inclinations and for faster azimuthal velocities. Besides, they also scale with the hot spot’s orbital radius, which is however antagonist to the magnitude of the  angular velocity \eqref{eq:Omega}. These  effects occur because radiation received at a given observation time is emitted at different retarded times along the hot spot’s orbit, effectively mixing different stages of its motion into a single observed frame. Specifically, the hot spot is elongated along its projected direction of motion when moving along an increasing time-of-flight gradient (see the approaching left side of the screen in Figure \hyperref[fig:starting_point]{\ref{fig:starting_point}}), and compressed along that same direction when moving along a decreasing time-of-flight gradient (see the receding right side of the screen in Figure \hyperref[fig:starting_point]{\ref{fig:starting_point}}).\footnote{For anticlockwise rotation, the pattern would be reversed, with elongation toward the right side and compression toward the left side.} This constitutes a generalisation of the Lampa-Terrell-Penrose effect \citep{Lampa1924SR,Terrell1959SR,Penrose1959SR} to non-rectilinear trajectories. Its analytical treatment is not essential for the aim of this paper, since incorporating it numerically into the analytical framework already developed provides the necessary cross-check and physical intuition on the analysed observables. We therefore refer to this hybrid analytical-numerical approach as the `semi-analytical' model throughout. Finally, even without an analytical form for the projected area of the emitter on the observer's screen, $P$, this quantity can be decomposed, without loss of generality, into static and time-varying components following the notation introduced in \eqref{eq:notation}:
\begin{align}
	P=P_0+P_t\:.
\end{align}

\subsection{Impacts on the light curve and its power spectrum}
\begin{figure*}[ht!]
	\centering
	\resizebox{14cm}{!}
	{\includegraphics{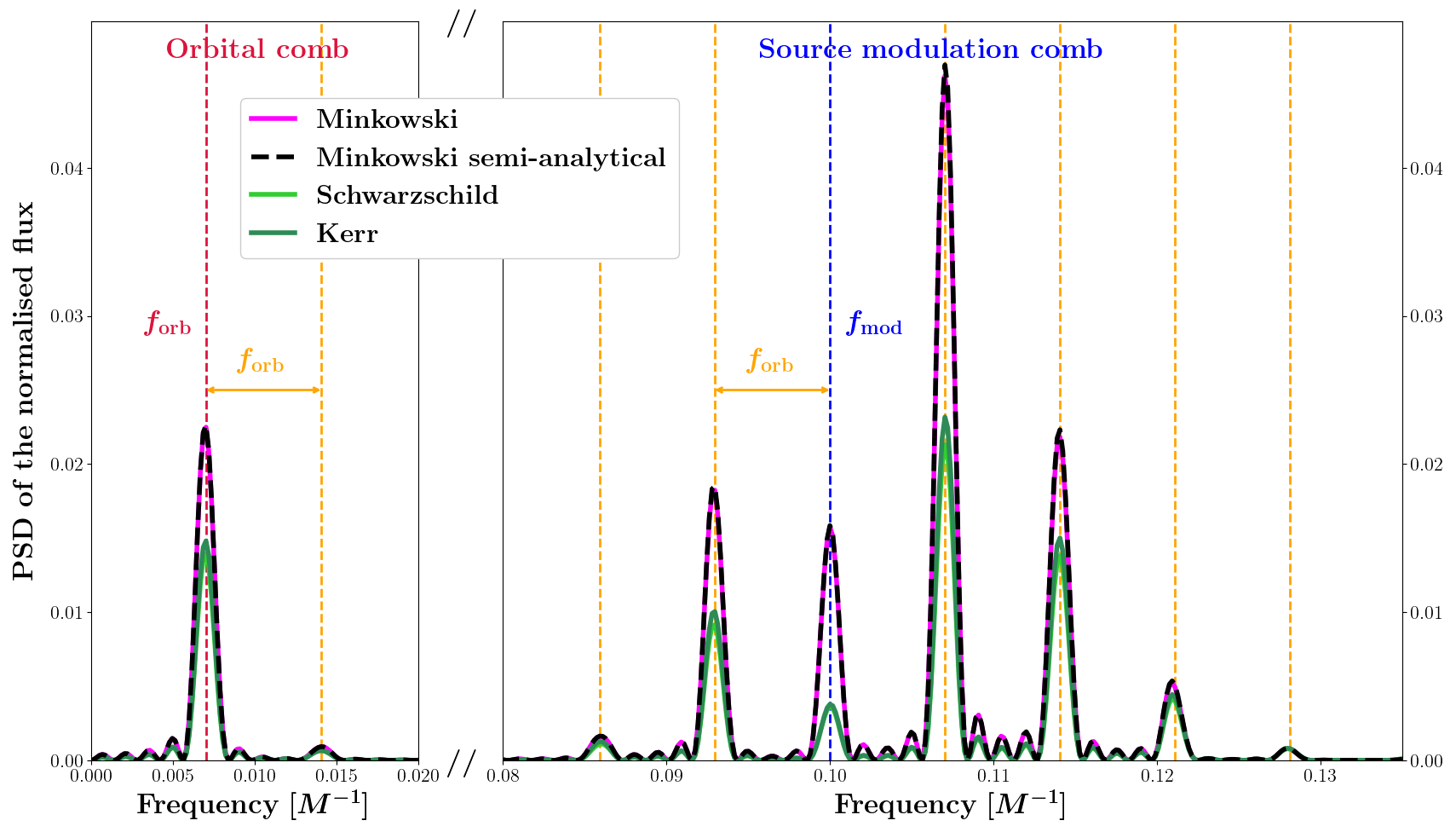}}
	\caption{Estimated power spectral densities of full-screen normalised light curves (see left panel of the fourth row of Figure \hyperref[fig:Explanatory]{\ref{fig:Explanatory}}), for an inclination $i=160^\circ$ in Minkowski (magenta), Schwarzschild (light green), and Kerr ($a=0.94$, dark green) space-times. Within flat space-time, the dashed black curve represents the estimated PSD of the semi-analytical prediction \eqref{eq:flux}, which perfectly matches the numerics. The two main peak families arise, in order, from relativistic beaming combined with variations of the apparent hot spot size on the screen, and from intrinsic modulation of the source. The corresponding frequency locations, indicated by dashed vertical lines, are exactly those expected from these mechanisms and their amplitude structure can be predicted by decomposing the different contributions as in equation \eqref{eq:flux_decomposed}.}
	\label{fig:PSDs}
\end{figure*}
To detect and characterise the periodic content of the light curves, we estimated their power spectral density (PSD), defined as the squared modulus of the Fast Fourier Transform (FFT) \citep{VanderPlas2018LombScargle}. Before computing the FFTs, we detrended the signals and padded them to a length of $2^{\lceil \log_2 (8\, n) \rceil}$ zeros, where $n$ is the number of time samples,\footnote{This optimises the efficiency of radix-2 FFT algorithms.} and we normalised the resulting transforms by a factor $2/n$. If $\mathcal{T}=n_\mathrm{orb}T_\mathrm{orb}$ is the total duration of the simulation and $\delta t=\mathcal{T}/n$ the frame cadence, the resolution in the frequency domain, $\Delta f \equiv 1/\mathcal{T}$, corresponds to the orbital frequency divided by the number of orbits. A detectable signal frequency must also lie well below the Nyquist frequency, $f_\mathrm{Nyq}\equiv1/2\delta t$.\footnote{For the values of Table \hyperref[tab:simu]{\ref{tab:simu}}, the Nyquist frequency is about 0.71 $M^{-1}$.} This condition is satisfied by $f_\mathrm{mod}$. With the normalisation convention that we adopted, a pure sinusoid $A\sin(2\pi f t)$ produces a peak at frequency $f$ and with power $A^2$ in the periodogram.\\
\indent Assuming that the redshift and the emitted intensity variation over the hotspot's spatial extent are negligible, we can evaluate these quantities at the geometric centre of the hot spot. According to the previous sections, the observed flux, $F$, results then from the product of the beaming factor, $G^\mathrm{\,c}\equiv(g^\mathrm{c})^{\,3}$, the emitted intensity, $I^\mathrm{\,em,\,c}$, and the projected area, $P$:
\begin{align}
	\mathscr{F}=G^\mathrm{\,c}I^\mathrm{\,em,\,c}P\:.\label{eq:flux}
\end{align}
By decomposing the factors of the product appearing in \eqref{eq:flux} into their zero-frequency and time-varying parts, as defined in \eqref{eq:notation}, the observed flux can be expanded into distinct contributions: 
\begin{align}
	\label{eq:flux_decomposed}
	\mathscr{F}=\: &G^\mathrm{\,c}_0\,I^\mathrm{\,em,\,c}_0P_0+\\
	&I^\mathrm{\,em,\,c}_0P_0\,G^\mathrm{\,c}_t+G^\mathrm{\,c}_0\,P_0\,I^\mathrm{\,em,\,c}_t+G^\mathrm{\,c}_0\,I^\mathrm{\,em,\,c}_0P_t+
	\notag\\
	& P_0\,G^\mathrm{\,c}_t\,I^\mathrm{\,em,\,c}_t +I^\mathrm{\,em,\,c}_0\,G^\mathrm{\,c}_t\,P_t+G^\mathrm{\,c}_0\,I^\mathrm{\,em,\,c}_tP_t+G^\mathrm{\,c}_t\,I^\mathrm{\,em,\,c}_tP_t\notag\:. \notag
\end{align}
The first line in \eqref{eq:flux_decomposed} represents a purely constant offset that contributes no power to the PSD, as the mean of the light curve was subtracted prior to spectral analysis. The second line isolates the contributions of the time-varying individual components, each scaled by a static factor, which are detailed sequentially in the following paragraphs. Finally, the third line collects the mixed terms arising from multi-component cross-modulation, which are discussed at the end of this section.\\
\indent The phase modulation of the observed light curve induced by light-travel-time or Rømer delays (see Section \hyperref[sec:times]{\ref{sec:times}}) causes the signal to undergo periodic temporal compression and stretching over one orbital period. This modulation of photon arrival times, which is visible in the three last rows of Figure \hyperref[fig:Explanatory]{\ref{fig:Explanatory}}, is also responsible for the Doppler shift in the electromagnetic frequency of the radiation.\footnote{Notably, Rømer conceived and explained the light-travel-time effect 166 years before Christian Doppler's 1842 treatise.} Although the associated shift in photon energy is not explicitly tracked in our analysis, its effects on the observed intensity are included, namely through relativistic beaming. In the periodogram, this harmonic behaviour manifests as a forest of sidebands, which we term the `source modulation comb', centred at $f_\mathrm{mod}$ (indicated by a blue dashed vertical line in Figure \hyperref[fig:PSDs]{\ref{fig:PSDs}}), separated by integer multiples of $f^\mathrm{\,c}_\mathrm{orb}$ (see orange dashed vertical lines in the right box of Figure \hyperref[fig:PSDs]{\ref{fig:PSDs}}). For the result \eqref{eq:Iem} in Minkowski space-time, for each discrete frequency $f_k=f_\mathrm{mod}+mf_\mathrm{orb}^\mathrm{\,c}$, the amplitudes of the corresponding multi-index coefficients that yield $m_k=m$ must be summed coherently. According to the expansion \eqref{eq:Iem}, the total PSD amplitudes are then directly given by $|\sum_{\boldsymbol{k}|m_k=m}A_k|^2$.\\
\indent The beaming effect, arising from the self-phase modulated redshift factor, causes the collected flux to decrease and increase periodically over one orbital period (see second row in Figure \hyperref[fig:Explanatory]{\ref{fig:Explanatory}}). In the periodogram, this contributes to a second forest of peaks, dubbed the `orbital comb', at the fundamental frequency $f_\mathrm{orb}^\mathrm{\,c}$ (indicated by a red dashed vertical line in Figure \hyperref[fig:PSDs]{\ref{fig:PSDs}}) and at its integer multiples (see orange dashed vertical lines in the left box of Figure \hyperref[fig:PSDs]{\ref{fig:PSDs}}). In this case, for the result \eqref{eq:beaming} in flat space-time, the individual amplitudes are not simply linear reflections of a single mode. As mixed frequency components originate from the expansion of higher-order powers of sinusoidal functions into sums, the amplitudes must be tracked carefully via their respective combined coefficients. For instance, for a second-power expansion, these paired interactions contain amplitude contributions of the form $|C_2^2\bar{A}_{k}\bar{A}_{k^\prime}/2|^2$ at the combined sum and difference frequencies, that is, at $f_{kk^\prime,+}=(m_k+m_k^\prime)f^\mathrm{\,c}_\mathrm{orb}$ and $f_{kk^\prime,-}=(m_k-m_k^\prime)f^\mathrm{\,c}_\mathrm{orb}$, respectively. Furthermore, exploiting the even symmetry of the cosine function, the negative frequency contributions can be mapped directly onto the positive frequency domain before estimating the single-sided PSD.\\
\indent The projected area of the hot spot also varies periodically over one orbital period (see Figure \hyperref[fig:terrell]{\ref{fig:terrell}}). Accordingly, its PSD exhibits a fundamental peak at $f_\mathrm{orb}^\mathrm{\,c}$ together with its harmonics, reflecting the increasingly non-sinusoidal nature of the evolution of the
projected area at higher inclinations (see Figure \hyperref[fig:terrell]{\ref{fig:terrell}}). Thus it contributes, alongside relativistic beaming, to the orbital comb.\\
\indent Apart from the combs of peaks described above, additional spectral features arise from the interaction between the different modulations.  These interactions generate a dense network of sum and difference frequencies through the product-to-sum identities of the underlying sinusoids, following a mechanism similar to the one described previously for the redshift. Crucially, these cross-modulation effects alter the amplitudes of the final peaks by distributing power among overlapping channels, but they do not introduce new spectral locations beyond the positions already established by the fundamental frequencies. As before, the symmetry properties of sine and cosine functions allow one to fold the spectrum onto positive frequencies when estimating the single-sided PSD amplitudes.

\section{Light curves analysis in curved space-time}
\label{sec:4}
Below, in Section \hyperref[sec:curvature]{\ref{sec:curvature}} we discuss how space-time curvature affects the various features of the light curve understood analytically within the Minkowski framework of Section \hyperref[sec:propagation]{\ref{sec:propagation}}. In the subsequent sections, we examine how these features can be used to conduct scientific analyses of black hole environments.
\begin{figure*}[ht!]
	\centering
	\resizebox{12.5cm}{!}
	{\includegraphics{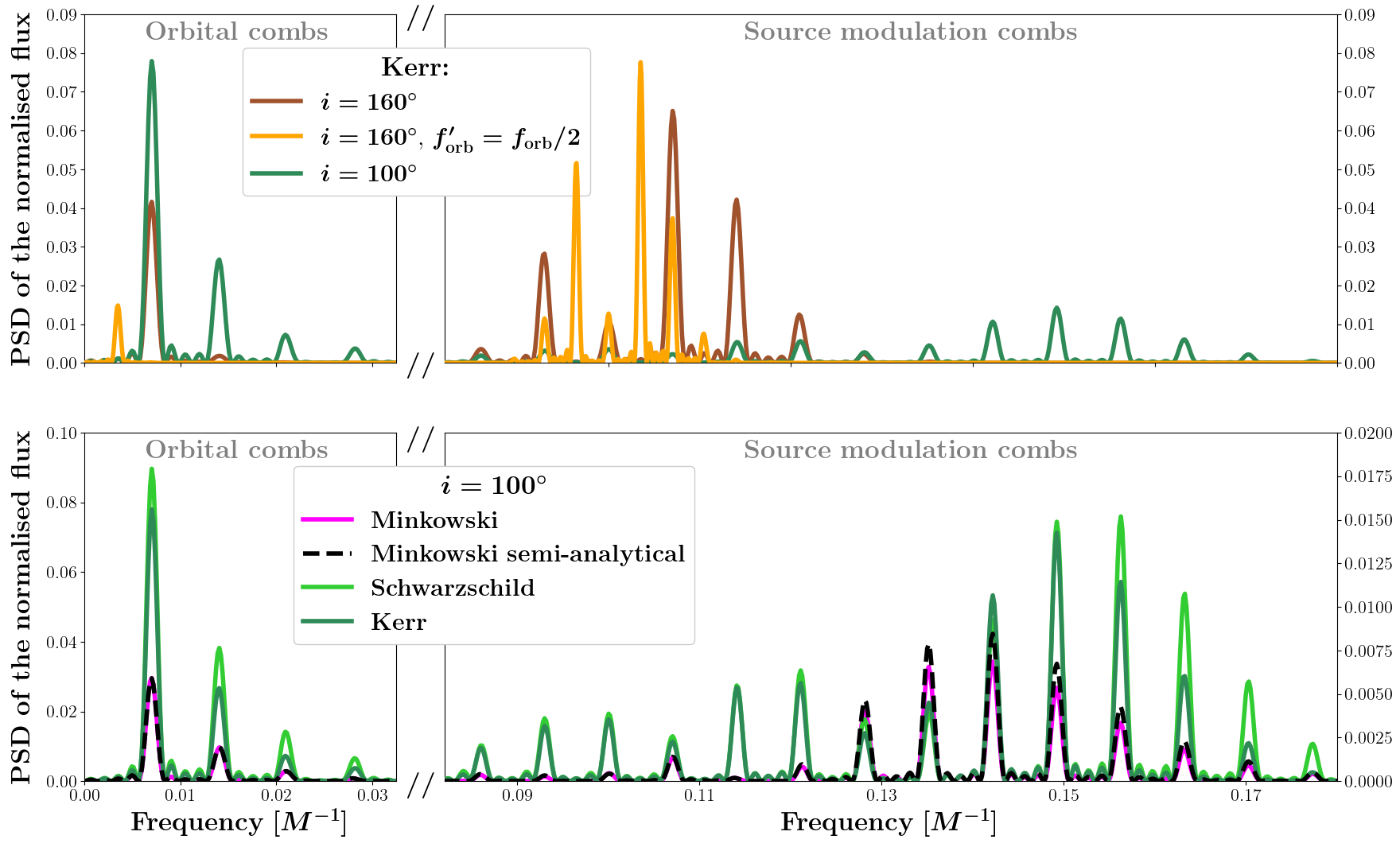}}
	\caption{Top panel: Estimated power spectral density of the full-screen normalised light curve in Kerr space-time for $i=160^\circ$ (brown and orange) or $i=100^\circ$ (dark green). The orange spectrum corresponds to a hot spot orbiting at half the base angular velocity \eqref{eq:Omega}. Higher inclination and orbital speed produce broader and blueshifted spectral envelopes, with the orbital frequency also setting the peak spacing. Bottom panel: Same as the top panel, but for an inclination $i=100^\circ$ in Minkowski (magenta), Schwarzschild (light green), and Kerr (a = 0.94, dark green) space-times. Note the different scales on the $y$-axes. At high inclination, differences between space-times extend beyond an overall rescaling of the spectrum.}
	\label{fig:dependence}
\end{figure*}

\subsection{Space-time curvature impact on light propagation effects}
\label{sec:curvature}
	In curved space-times, the time of flight, $F$, between the hot spot and the observer, introduced in equation \eqref{eq:time}, is no longer determined by the simple Euclidean distance. Instead, one must account for gravitational lensing effects, which make light-travel time increase near the black hole as photons undergo significant angular deflection to reach infinity. Furthermore, in the case of a rotating black hole, Lense-Thirring frame-dragging effects alter the azimuthal position of the hot spot in the equatorial plane, thereby modifying the corresponding time of flight.
	A visualisation of the emission coordinates in different metrics is provided in Appendix \hyperref[app:hit]{\ref{app:hit}}. Given the analytical relations between the observer's screen and the impact coordinates in the Kerr metric \citep{CarterPRD,Bardeen1973,Viergutz1993A&A,RauchBlandford1994ApJ,DexterAgol1994ApJ,Gralla2020Lensing}, we expect the lensing transfer function $F$ to be derivable analytically. Provided that this mapping is infinitely derivable, an analytical solution for $t^\mathrm{em}(t^\mathrm{obs})$ could be obtained via a series expansion similar to the one performed for the Minkowski case (see Appendix \hyperref[app:Minko]{\ref{app:Minko}}). A formal treatment in the Kerr metric is however omitted here, both because the Minkowski limit, which is mathematically much simpler, already captures the fundamental mechanism behind the light curve phase-modulation, providing the necessary physical intuition to interpret the general-relativistic simulations, and because such a refined treatment would be disproportionate to the simplified setup considered in this study. In particular, as shown in Figure \hyperref[fig:inversion]{\ref{fig:inversion}}, the relations between emission and observation times in all three metrics are nearly indistinguishable when the source is at its closest approach to the observer, that is, when photons undergo minimal gravitational deflection, thus closely following Minkowski-like trajectories. In the same Figure  \hyperref[fig:inversion]{\ref{fig:inversion}}, notice however that curved space-times exhibit a non-equal distribution of observed time spent between the top and bottom image quadrants. Specifically, the source appears to spend more time in the bottom-right quadrant compared to the Minkowski case and this behaviour is more pronounced in the Schwarzschild metric than in the Kerr metric.\\
	\indent The redshift factor \eqref{eq:redshift} is also expected to be analytically computable in the Kerr metric, as the photon four-momentum along the geodesic is known in closed form \citep{Gralla2020Lensing,Cardenas2023AART}. Given an analytical expression for $t^\mathrm{em}(t^\mathrm{obs})$, one could explicitly formulate $g(t^\mathrm{obs})$ to describe the modulated flux of the hot spot, but for the same reasons as before, we restrict our current analysis to the numerical results presented in Figure \hyperref[fig:redshift]{\ref{fig:redshift}}. We notice that while the beaming effect in curved space-times remains periodic over orbital periods, it lacks the symmetry around the maximum seen in the Minkowski case. This asymmetry is the consequence of the variation of the emitted photon four-momentum as the spot moves from the closest part to the furthest part of the orbit and leads to modification of amplitudes in the periodogram. Furthermore, Figure \hyperref[fig:redshift]{\ref{fig:redshift}} illustrates that the redshift factor in curved space-times is generally smaller than in the Minkowski limit and this remains true at smaller inclination (see Figure \hyperref[fig:g160]{\ref{fig:g160}}). This is due to the significant gravitational redshift imposed by the central mass, in addition to the dynamical redshift contribution, which leads to a systematic reduction in the overall observed flux and, by consequence, power in the PSD peaks.\\
	\indent In curved space-time, the top-bottom symmetry is also broken for the projected area of the source (see Figure \hyperref[fig:terrell]{\ref{fig:terrell}}). Gravitational lensing leads to a larger projected area of the hot spot on the far side of the orbit compared to its closest approach. This purely general relativistic magnification effect further contributes to the asymmetry of the observed light curve and the subsequent distribution of power in the periodogram's sidebands.
	
	\subsection{Spectral variations of full-image light curves across astrophysical and geometrical configurations}
	As established above, the structure of the light-curve PSD is governed by a combination of the orbital setup, the hot spot properties, and the space-time geometry. We now describe the qualitative dependence of the spectral features on these factors.\\
	\indent For a given space-time, increasing the inclination leads to a stronger contrast in amplitude between the orbital and source modulation combs, as well as to their systematic broadening and shift toward higher frequencies, as visible when comparing the brown PSD at $i=160^\circ$ to the green PSD at $100^\circ$ in the top panel of Figure \hyperref[fig:dependence]{\ref{fig:dependence}}. The source velocity produces similar effects, as decreasing it increases the contrast, while reversing the relative amplitude hierarchy, shifts the spectrum toward lower frequencies, and reduces the width of the envelopes. Additionally, it modifies the separation of the individual peaks, which is set by the orbital angular frequency. This is illustrated in the top panel of Figure \hyperref[fig:dependence]{\ref{fig:dependence}} by the comparison between the brown PSD at $i=160^\circ$ and the orange PSD obtained at the same inclination but with half the velocity. We stress that all these astrophysical effects are captured by the semi-analytical model \eqref{eq:flux} in Minkowski space-time, which therefore provides a valuable first-order tool for interpreting the observed spectral structure, although the dependence of the phase-modulation coefficients discussed in Sections \hyperref[sec:times]{\ref{sec:times}} and \hyperref[sec:beam]{\ref{sec:beam}} on the inclination and angular velocity is not straightforward to guess intuitively.\\
	\indent For the same astrophysical configuration, the distinctions between space-times become more pronounced at high inclination, as can be seen by comparing the results at $i=160^\circ$, shown in Figure \hyperref[fig:PSDs]{\ref{fig:PSDs}}, with the power spectra in the bottom panel of Figure \hyperref[fig:dependence]{\ref{fig:dependence}}, obtained from light curves at $i=100^\circ$ (see Appendix \hyperref[app:additional]{\ref{app:additional}} for more details on the high-inclination simulations). At low inclination, the Kerr and Schwarzschild cases are nearly identical, with only minor differences with respect to Minkowski space-time once an overall scaling is taken into account. At high inclination, by contrast, the power spectra in curved space-time appear blueshifted, and their orbital combs exhibit a substantial enhancement in harmonic power. This enhancement results from the larger frequency range spanned by the projected-area variations and is slightly more pronounced in Schwarzschild space-time (see the bottom right panel of Figure \hyperref[fig:terrell]{\ref{fig:terrell}}). We also note that this same contribution accounts for the apparent broadening of the low-frequency tail of the source modulation comb in curved space-time, rather than reflecting an intrinsic broadening of the modulation envelope itself, as confirmed by simulations at a higher modulation frequency.
	
	\subsection{Doppler-induced diagnostics from relative estimators of two-angular-sector image power spectra}
	\begin{figure}[ht!]
		\centering
		\includegraphics[width=9.5cm]{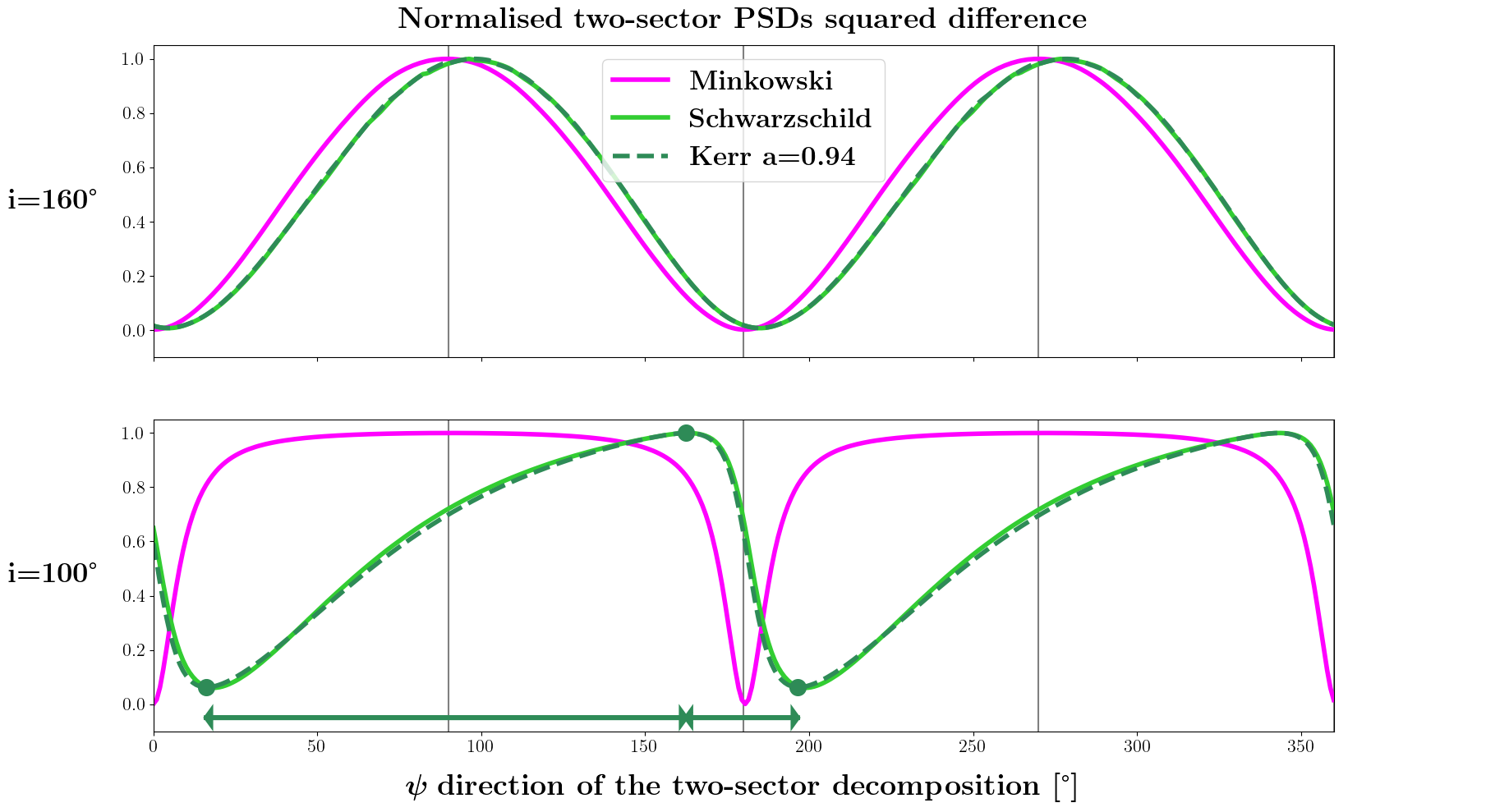}
		\caption{Squared difference between the two-sector light-curve periodograms at $i=160^\circ$ (top panel) and $i=100^\circ$ (bottom panel). Profiles are shown for Minkowski (magenta), Schwarzschild (light green), and Kerr ($a=0.94$, dark green) space-times. High values corresponds to the orientation where the Doppler shift is most pronounced. This direction coincides with the projected angular momentum of the source in Minkowski space-time (vertical lines mark integer multiples of $90^\circ$), but increasingly deviates from it with inclination in curved space-times. For the Kerr profiles in the bottom panel, arrows indicate the separation between successive extrema marked by dots.}
		\label{fig:orientation}
	\end{figure}
\begin{figure}[ht!]
		\centering
		\includegraphics[width=9.5cm]{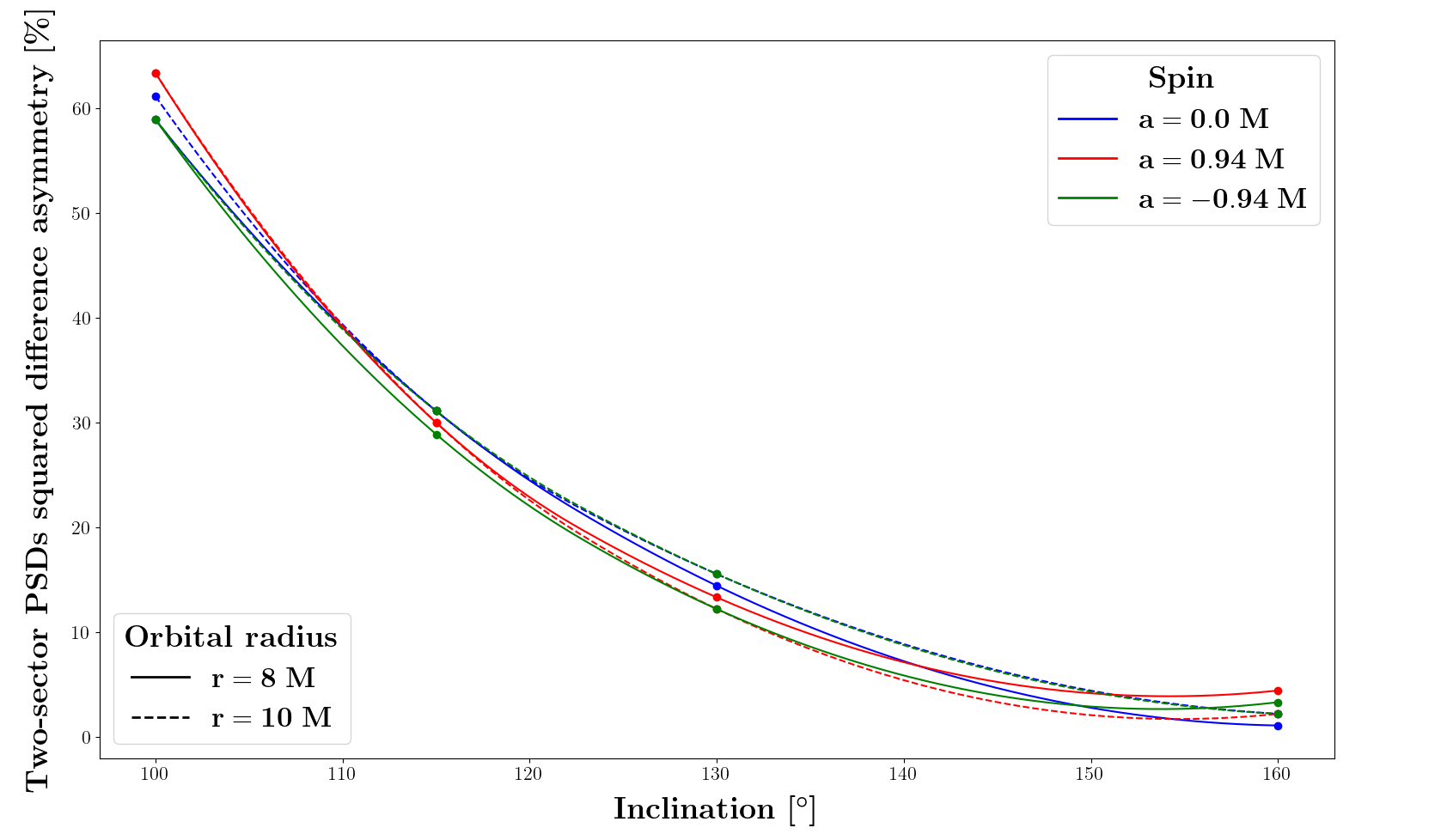}
		\caption{Asymmetry in the squared difference between the two-sector periodograms in curved space-times, shown in green in Figure \hyperref[fig:orientation]{\ref{fig:orientation}}, as a function of inclination, and for various astrophysical configurations, which have only a minor impact on the overall trend. Different line styles denote different orbital radii of the hot spot, while colours distinguish different black hole spins. Dots indicate measured values, and lines show quadratic polynomial interpolations. This asymmetry therefore provides a diagnostic of the system's inclination.}
		\label{fig:asymmetry}
	\end{figure}

	Previously, we considered the integrated light curve over the full screen, whose characterisation results from the evolution of the signal across different regions of the image plane. We now investigate what additional information can be obtained by isolating the light curves from fixed screen angular sectors and comparing their individual spectral signatures. In particular, we considered the two halves of the image defined with respect to an axis lying along an arbitrary polar direction $\psi$, passing from the barycentre of the traced trajectory. Following the discussion of Section \hyperref[sec:propagation]{\ref{sec:propagation}}, the associated light curves probe different portions of the phase-modulated oscillatory signal, potentially leading to redshift or blueshift of the spectral frequencies (see last row of Figure \hyperref[fig:Explanatory]{\ref{fig:Explanatory}}). To retrieve information on the system through the expected Doppler shift, we therefore scanned over all possible $\psi$-directions and took as an estimator the squared difference between the sector periodograms, summed over all frequency bins.\\
	\indent At low inclination, as expected from intuition, the difference between the spectra is maximised when the cut direction of the decomposition matches that of the projected angular momentum vector of the hot spot, isolating then the approaching and receding, or equivalently blueshifted and redshifted, sides of the orbit. This is illustrated in the top panel of Figure \hyperref[fig:orientation]{\ref{fig:orientation}}, where the maximum spectral difference is reached at $\psi=90^\circ$, matching the prescribed orientation of our simulations. We note that while the correct value is exactly retrieved in the Minkowski case, a slight deviation is present in Schwarzschild and Kerr space-times, due to the lensing effects detailed below.\\
	\indent At high inclination, distinct trends emerge in flat and curved space-times. In flat space-time, we recover the behaviour described above, with a more squared and less Gaussian shape, consistent with the limiting edge-on case, in which the projected orbit is a horizontal line. In this limit, all cuts yield the same maximum value, except for $\psi=0^\circ$, for which the spectral difference vanishes.
	In curved space-times, instead, the profiles appear distorted, with neither a clear symmetry nor a maximum around the expected projected angular momentum vector of the source, and no zero minimum in the perpendicular direction. Both effects arise from the lensing-induced differences between the near and far sides of the orbit, as discussed in Section \hyperref[sec:curvature]{\ref{sec:curvature}}, preventing equal contributions from the two half sectors of the screen and introducing significant asymmetries in the projected area.\\
	\indent These distinct behaviours can be quantified by measuring the asymmetry of the curves shown in Figure \hyperref[fig:orientation]{\ref{fig:orientation}}, defined as the difference between the separations of the first and second pairs of maxima, as indicated by the arrows in the figure. In flat space-time, the asymmetry vanishes at all inclinations, whereas in curved space-time it increases with inclination, following a trend that is insensitive to the hot spot's orbital radius and the black hole spin, as illustrated in Figure  \hyperref[fig:asymmetry]{\ref{fig:asymmetry}}. Differently stated, the non-zero magnitude of this new observable probes space-time curvature and can be used to constrain inclination.\footnote{For Sgr A*, the inclination has been inferred with observational uncertainties and under specific assumptions \citep{GRAVITY2018Flares}, whereas it may be unknown for other sources targeted by future, higher-angular-resolution observations.} Based on Figure \hyperref[fig:orientation]{\ref{fig:orientation}}, when the retrieved inclination is low, the location of the maximum asymmetry provides the additional information needed to determine the orientation of the system.
	
	\section{Discussion}
	\label{sec:end} 
	
	\subsection{Conclusion and limitations}
    In this article, we have proposed a novel approach to extract astrophysical and geometrical information on black hole environments based on the observed variability of the direct emission from the accreting matter. Our focus in this first study has been on power spectral densities from resolved direct light curves of a single fluctuating hot spot, a structure that naturally forms through magnetic reconnection in fluid and kinetic simulations \citep{Ripperda2020HotSpot,Mehlhaff2025PIC} and is often invoked as a key source of variability. Our approach is relevant to flares observed by GRAVITY \citep{GRAVITY2018Flares} and ALMA \citep{Wielgus2022Flare}, as well as to future black hole movies from extensions of the EHT \citep{Johnson2023ngEHT,Johnson2024BHEX}. The contributions of this work are two-fold. First, we have provided a physical interpretation of the periodic content expected in light curves generated by a source on a circular orbit and with emission modulated at a single intrinsic frequency, through a semi-analytical model of light-propagation effects in flat space-time, reproducing all qualitative features observed in curved space-time at moderate inclination. Second, we have explored how such spectra can be used for scientific inference. On the one hand, we have identified the dependence of full-image spectra on observer inclination, source velocity, and space-time geometry. On the other hand, we have shown how two-sector screen decompositions can highlight Doppler-induced effects, which reveal lensing asymmetries, and can be used to constrain the inclination and orientation of the system.\\
    \indent A series of assumptions and approximations make this preliminary paper mainly a conceptual investigation. Further developments in the modelling will be required to bring the approach closer to practical applicability to data. First, we considered time series spanning five orbits with very high temporal resolution, whereas hot-spot observations typically cover at best only one orbit resolved over timescales of minutes \citep{GRAVITY2018Flares}. In this regime, the PSD would be contaminated by the signal of the restricted observational window, making individual peaks difficult to distinguish, although the global position and shape of the spectral envelopes would remain informative, and the frequency range accessible to the analysis would be narrower. Furthermore, although the emission modulation frequency of the hot spot was chosen within the range found in GRMHD and GRPIC simulations \citep{Porth2019GRMHD,Mehlhaff2025PIC}, the underlying emitted intensity is not tied to any specific physical emission mechanism, and we adopted a simplified prescription for proof-of-principle purposes. Finally, we neglected the disc background and adopted a hot-spot modulation whose maximum amplitude matches its base value. Exploring more realistic values for both components, thereby clarifying the role of their relative amplitudes, lies beyond the scope of this work. The primary objective of this toy model is to assess the validity of our approach in a simplified setting and a more comprehensive investigation is left for future work.
	
	\subsection{Perspectives}
	While the present framework serves as a foundational step, it provides a groundwork for more complex and realistic scenarios. We outline below the immediate and natural extensions of this project that we intend to pursue.\\
	\indent As a first step in understanding a fluctuating accretion flow, it would be worth investigating the increased intricacy that arises when transitioning from a single variable emitter to multiple hot spots sharing the same intrinsic modulation frequency. 
	Future work should determine whether specific configuration details remain identifiable, potentially through statistical analysis.\\
	\indent Then, although this study isolates a single characteristic modulation frequency, more sophisticated numerical simulations reveal a full turbulent cascade spanning a wide range of frequencies \citep{Porth2019GRMHD,Mehlhaff2025PIC}. Future developments of our model should incorporate a continuous power-law spectrum of frequencies to capture the realistic multi-scale fluctuations of the disc, rather than monochromatic drivers. This will require developing or adopting analytically tractable stochastic disc models, such as that of \citet{Lee2021Inhomogeneities}, or physically motivated semi-analytical constructions that capture key temporal properties of plasma simulations, as those considered in \citet{Mummery2026TDT}.\\
	\indent An important next step will be to assess whether angular-sector decompositions are feasible with real interferometric data. If so, the promising results obtained with our initial two-sector cut, the most natural first choice, warrant further exploration of this spatially resolved methodology. Future investigations could explore finer decompositions, alternative image-partitioning schemes, and direct analyses of interferometric signals to maximise the diagnostic information.\\
	\indent Finally, while the present study relies on standard Fourier analysis, which implicitly assumes harmonic oscillations that persist globally over the entire time series, more advanced signal-processing techniques would be valuable. For instance, employing wavelet transforms would allow us to track transient, localised oscillations and capture the dynamic evolution of fluctuations across both spatial and temporal scales.
	
	\begin{acknowledgements}
	The authors sincerely thank Alejandro Cárdenas-Avendaño for valuable feedback.
	\end{acknowledgements}
	
	\bibliographystyle{aa}
	\bibliography{Bibliography}

@ARTICLE{Balick1974SgrA,
	author = {{Balick}, B. and {Brown}, R.~L.},
	title = "{Intense sub-arcsecond structure in the galactic center.}",
	journal = {\apj},
	year = 1974,
	month = dec,
	volume = {194},
	pages = {265-270},
	doi = {10.1086/153242},
	adsurl = {https://ui.adsabs.harvard.edu/abs/1974ApJ...194..265B}
}

@ARTICLE{RauchBlandford1994ApJ,
       author = {{Rauch}, Kevin P. and {Blandford}, Roger D.},
        title = "{Optical Caustics in a Kerr Spacetime and the Origin of Rapid X-Ray Variability in Active Galactic Nuclei}",
      journal = {\apj},
         year = 1994,
        month = jan,
       volume = {421},
        pages = {46},
          doi = {10.1086/173625},
       adsurl = {https://ui.adsabs.harvard.edu/abs/1994ApJ...421...46R}
}

@ARTICLE{Do2019S2,
	author = {{Do}, Tuan and {Hees}, Aurelien and {Ghez}, Andrea and {Martinez}, Gregory D. and {Chu}},
	title = "{Relativistic redshift of the star S0-2 orbiting the Galactic Center supermassive black hole}",
	journal = {\science},
	year = 2019,
	month = aug,
	volume = {365},
	number = {6454},
	pages = {664-668},
	doi = {10.1126/science.aav8137},
	archivePrefix = {arXiv},
	eprint = {1907.10731},
	primaryClass = {astro-ph.GA},
	adsurl = {https://ui.adsabs.harvard.edu/abs/2019Sci...365..664D}
}

@ARTICLE{Narayan1995SgrA,
	author = {{Narayan}, Ramesh and {Yi}, Insu and {Mahadevan}, Rohan},
	title = "{Explaining the spectrum of Sagittarius A$^{*}$ with a model of an accreting black hole}",
	journal = {\nat},
	year = 1995,
	month = apr,
	volume = {374},
	number = {6523},
	pages = {623-625},
	doi = {10.1038/374623a0},
	adsurl = {https://ui.adsabs.harvard.edu/abs/1995Natur.374..623N}
}

@ARTICLE{Yuan2003RIAF,
	author = {{Yuan}, Feng and {Quataert}, Eliot and {Narayan}, Ramesh},
	title = "{Nonthermal Electrons in Radiatively Inefficient Accretion Flow Models of Sagittarius A*}",
	journal = {\apj},
	year = 2003,
	month = nov,
	volume = {598},
	number = {1},
	pages = {301-312},
	doi = {10.1086/378716},
	archivePrefix = {arXiv},
	eprint = {astro-ph/0304125},
	primaryClass = {astro-ph},
	adsurl = {https://ui.adsabs.harvard.edu/abs/2003ApJ...598..301Y}
}

@ARTICLE{Yuan2014ARIAF,
	author = {{Yuan}, Feng and {Narayan}, Ramesh},
	title = "{Hot Accretion Flows Around Black Holes}",
	journal = {\araa},
	year = 2014,
	month = aug,
	volume = {52},
	pages = {529-588},
	doi = {10.1146/annurev-astro-082812-141003},
	archivePrefix = {arXiv},
	eprint = {1401.0586},
	primaryClass = {astro-ph.HE},
	adsurl = {https://ui.adsabs.harvard.edu/abs/2014ARA&A..52..529Y}
}

@ARTICLE{GRAVITY2019Distance,
	author = {{GRAVITY Collaboration}},
	title = "{A geometric distance measurement to the Galactic center black hole with 0.3\% uncertainty}",
	journal = {\aap},
	year = 2019,
	month = may,
	volume = {625},
	eid = {L10},
	pages = {L10},
	doi = {10.1051/0004-6361/201935656},
	archivePrefix = {arXiv},
	eprint = {1904.05721},
	primaryClass = {astro-ph.GA},
	adsurl = {https://ui.adsabs.harvard.edu/abs/2019A&A...625L..10G}
}

@ARTICLE{Brown1982Radio,
	author = {{Brown}, R.~L. and {Lo}, K.~Y.},
	title = "{Variability of the compact radio source at the Galactic Center.}",
	journal = {\apj},
	year = 1982,
	month = feb,
	volume = {253},
	pages = {108-114},
	doi = {10.1086/159615},
	adsurl = {https://ui.adsabs.harvard.edu/abs/1982ApJ...253..108B}
}

@ARTICLE{Baganoff2001Xray,
	author = {{Baganoff}, F.~K. and {Bautz}, M.~W. and {Brandt}, W.~N. and {Chartas}, G. and {Feigelson}, E.~D. and {Garmire}, G.~P. and {Maeda}, Y. and {Morris}, M. and {Ricker}, G.~R. and {Townsley}, L.~K. and {Walter}, F.},
	title = "{Rapid X-ray flaring from the direction of the supermassive black hole at the Galactic Centre}",
	journal = {\nat},
	year = 2001,
	month = sep,
	volume = {413},
	number = {6851},
	pages = {45-48},
	doi = {10.1038/35092510},
	archivePrefix = {arXiv},
	eprint = {astro-ph/0109367},
	primaryClass = {astro-ph},
	adsurl = {https://ui.adsabs.harvard.edu/abs/2001Natur.413...45B}
}

@ARTICLE{Ghenzel2003NIR,
	author = {{Genzel}, R. and {Sch{\"o}del}, R. and {Ott}, T. and {Eckart}, A. and {Alexander}, T. and {Lacombe}, F. and {Rouan}, D. and {Aschenbach}, B.},
	title = "{Near-infrared flares from accreting gas around the supermassive black hole at the Galactic Centre}",
	journal = {\nat},
	year = 2003,
	month = oct,
	volume = {425},
	number = {6961},
	pages = {934-937},
	doi = {10.1038/nature02065},
	archivePrefix = {arXiv},
	eprint = {astro-ph/0310821},
	primaryClass = {astro-ph},
	adsurl = {https://ui.adsabs.harvard.edu/abs/2003Natur.425..934G}
}

@ARTICLE{Ghez2004IR,
	author = {{Ghez}, A.~M. and {Wright}, S.~A. and {Matthews}, K. and {Thompson}, D. and {Le Mignant}, D. and {Tanner}, A. and {Hornstein}, S.~D. and {Morris}, M. and {Becklin}, E.~E. and {Soifer}, B.~T.},
	title = "{Variable Infrared Emission from the Supermassive Black Hole at the Center of the Milky Way}",
	journal = {\apjl},
	year = 2004,
	month = feb,
	volume = {601},
	number = {2},
	pages = {L159-L162},
	doi = {10.1086/382024},
	archivePrefix = {arXiv},
	eprint = {astro-ph/0309076},
	primaryClass = {astro-ph},
	adsurl = {https://ui.adsabs.harvard.edu/abs/2004ApJ...601L.159G}
}

@ARTICLE{Do2009NIR,
	author = {{Do}, T. and {Ghez}, A.~M. and {Morris}, M.~R. and {Yelda}, S. and {Meyer}, L. and {Lu}, J.~R. and {Hornstein}, S.~D. and {Matthews}, K.},
	title = "{A Near-Infrared Variability Study of the Galactic Black Hole: A Red Noise Source with NO Detected Periodicity}",
	journal = {\apj},
	year = 2009,
	month = feb,
	volume = {691},
	number = {2},
	pages = {1021-1034},
	doi = {10.1088/0004-637X/691/2/1021},
	archivePrefix = {arXiv},
	eprint = {0810.0446},
	primaryClass = {astro-ph},
	adsurl = {https://ui.adsabs.harvard.edu/abs/2009ApJ...691.1021D}
}

@ARTICLE{GRAVITY2018Flares,
	author = {{GRAVITY Collaboration}},
	title = "{Detection of orbital motions near the last stable circular orbit of the massive black hole SgrA*}",
	journal = {\aap},
	year = 2018,
	month = oct,
	volume = {618},
	eid = {L10},
	pages = {L10},
	doi = {10.1051/0004-6361/201834294},
	archivePrefix = {arXiv},
	eprint = {1810.12641},
	primaryClass = {astro-ph.GA},
	adsurl = {https://ui.adsabs.harvard.edu/abs/2018A&A...618L..10G}
}

@ARTICLE{Wielgus2022Flare,
	author = {{Wielgus}, M. and {Moscibrodzka}, M. and {Vos}, J. and {Gelles}, Z. and {Mart{\'\i}-Vidal}, I. and {Farah}, J. and {Marchili}, N. and {Goddi}, C. and {Messias}, H.},
	title = "{Orbital motion near Sagittarius A$^{*}$ . Constraints from polarimetric ALMA observations}",
	journal = {\aap},
	year = 2022,
	month = sep,
	volume = {665},
	eid = {L6},
	pages = {L6},
	doi = {10.1051/0004-6361/202244493},
	archivePrefix = {arXiv},
	eprint = {2209.09926},
	primaryClass = {astro-ph.HE},
	adsurl = {https://ui.adsabs.harvard.edu/abs/2022A&A...665L...6W}
}

@ARTICLE{EHTISgrAShadow,
	author = {{Event Horizon Telescope Collaboration}},
	title = "{First Sagittarius A* Event Horizon Telescope Results. I. The Shadow of the Supermassive Black Hole in the Center of the Milky Way}",
	journal = {\apjl},
	year = 2022,
	month = may,
	volume = {930},
	number = {2},
	eid = {L12},
	pages = {L12},
	doi = {10.3847/2041-8213/ac6674},
	adsurl = {https://ui.adsabs.harvard.edu/abs/2022ApJ...930L..12E}
}

@ARTICLE{EHTIVSgrAVariability,
	author = {{Event Horizon Telescope Collaboration}},
	title = "{First Sagittarius A* Event Horizon Telescope Results. IV. Variability, Morphology, and Black Hole Mass}",
	journal = {\apjl},
	year = 2022,
	month = may,
	volume = {930},
	number = {2},
	eid = {L15},
	pages = {L15},
	doi = {10.3847/2041-8213/ac6736},
	archivePrefix = {arXiv},
	eprint = {2311.08697},
	primaryClass = {astro-ph.HE},
	adsurl = {https://ui.adsabs.harvard.edu/abs/2022ApJ...930L..15E}
}

@ARTICLE{EHTVSgrAModels,
	author = {{Event Horizon Telescope Collaboration}},
	title = "{First Sagittarius A* Event Horizon Telescope Results. V. Testing Astrophysical Models of the Galactic Center Black Hole}",
	journal = {\apjl},
	year = 2022,
	month = may,
	volume = {930},
	number = {2},
	eid = {L16},
	pages = {L16},
	doi = {10.3847/2041-8213/ac6672},
	adsurl = {https://ui.adsabs.harvard.edu/abs/2022ApJ...930L..16E}
}

@ARTICLE{EHTVIII2024Polarised,
	author = {{Event Horizon Telescope Collaboration}}, 
	title = "{First Sagittarius A* Event Horizon Telescope Results. VIII. Physical Interpretation of the Polarized Ring}",
	journal = {\apjl},
	year = 2024,
	month = apr,
	volume = {964},
	number = {2},
	eid = {L26},
	pages = {L26},
	doi = {10.3847/2041-8213/ad2df1},
	adsurl = {https://ui.adsabs.harvard.edu/abs/2024ApJ...964L..26E}
}

@ARTICLE{Wielgus2022Millimeter,
	author = {{Wielgus}, Maciek and {Marchili}, Nicola and {Mart{\'\i}-Vidal}, Iv{\'a}n and {Keating}, Garrett K. and {Ramakrishnan}},
	title = "{Millimeter Light Curves of Sagittarius A* Observed during the 2017 Event Horizon Telescope Campaign}",
	journal = {\apjl},
	year = 2025,
	month = jan,
	volume = {930},
	number = {2},
	eid = {L19},
	pages = {L19},
	doi = {10.3847/2041-8213/ac6428},
	archivePrefix = {arXiv},
	eprint = {2207.06829},
	primaryClass = {astro-ph.HE},
	adsurl = {https://ui.adsabs.harvard.edu/abs/2022ApJ...930L..19W}
}

@ARTICLE{Ressler2020StellarFeed,
	author = {{Ressler}, Sean M. and {White}, Christopher J. and {Quataert}, Eliot and {Stone}, James M.},
	title = "{Ab Initio Horizon-scale Simulations of Magnetically Arrested Accretion in Sagittarius A* Fed by Stellar Winds}",
	journal = {\apjl},
	year = 2020,
	month = jun,
	volume = {896},
	number = {1},
	eid = {L6},
	pages = {L6},
	doi = {10.3847/2041-8213/ab9532},
	archivePrefix = {arXiv},
	eprint = {2006.00005},
	primaryClass = {astro-ph.HE},
	adsurl = {https://ui.adsabs.harvard.edu/abs/2020ApJ...896L...6R}
}

@ARTICLE{Gezel2010SgrAOverview,
	author = {{Genzel}, Reinhard and {Eisenhauer}, Frank and {Gillessen}, Stefan},
	title = "{The Galactic Center massive black hole and nuclear star cluster}",
	journal = {\rmp},
	year = 2010,
	month = oct,
	volume = {82},
	number = {4},
	pages = {3121-3195},
	doi = {10.1103/RevModPhys.82.3121},
	archivePrefix = {arXiv},
	eprint = {1006.0064},
	primaryClass = {astro-ph.GA},
	adsurl = {https://ui.adsabs.harvard.edu/abs/2010RvMP...82.3121G}
}

@ARTICLE{DexterAgol1994ApJ,
       author = {{Rauch}, Kevin P. and {Blandford}, Roger D.},
        title = "{Optical Caustics in a Kerr Spacetime and the Origin of Rapid X-Ray Variability in Active Galactic Nuclei}",
      journal = {\apj},
         year = 1994,
        month = jan,
       volume = {421},
        pages = {46},
          doi = {10.1086/173625},
       adsurl = {https://ui.adsabs.harvard.edu/abs/1994ApJ...421...46R}
}

@article{CarterPRD,
  title = {Global Structure of the Kerr Family of Gravitational Fields},
  author = {Carter, Brandon},
  journal = {Phys. Rev.},
  volume = {174},
  issue = {5},
  pages = {1559--1571},
  numpages = {0},
  year = {1968},
  month = {Oct},
  publisher = {American Physical Society},
  doi = {10.1103/PhysRev.174.1559},
  url = {https://link.aps.org/doi/10.1103/PhysRev.174.1559}
}

@ARTICLE{Viergutz1993A&A,
       author = {{Viergutz}, S.~U.},
        title = "{Image generation in Kerr geometry. I. Analytical investigations on the stationary emitter-observer problem}",
      journal = {\aap},
         year = 1993,
        month = may,
       volume = {272},
        pages = {355},
       adsurl = {https://ui.adsabs.harvard.edu/abs/1993A&A...272..355V}
}

@INPROCEEDINGS{Bardeen1973,
       author = {{Bardeen}, J.~M.},
        title = "{Timelike and null geodesics in the Kerr metric.}",
    booktitle = {Black Holes (Les Astres Occlus)},
         year = 1973,
       editor = {{Dewitt}, C. and {Dewitt}, B.~S.},
        month = jan,
        pages = {215-239},
       adsurl = {https://ui.adsabs.harvard.edu/abs/1973blho.conf..215B}
}

@ARTICLE{Chan2009MRI,
	author = {{Chan}, Chi-kwan and {Liu}, Siming and {Fryer}, Christopher L. and {Psaltis}, Dimitrios and {{\"O}zel}, Feryal and {Rockefeller}, Gabriel and {Melia}, Fulvio},
	title = "{MHD Simulations of Accretion onto Sgr A*: Quiescent Fluctuations, Outbursts, and Quasiperiodicity}",
	journal = {\apj},
	year = 2009,
	month = aug,
	volume = {701},
	number = {1},
	pages = {521-534},
	doi = {10.1088/0004-637X/701/1/521},
	archivePrefix = {arXiv},
	eprint = {astro-ph/0611269},
	primaryClass = {astro-ph},
	adsurl = {https://ui.adsabs.harvard.edu/abs/2009ApJ...701..521C}
}

@ARTICLE{Dexter2013Shocks,
	author = {{Dexter}, Jason and {Fragile}, P. Chris},
	title = "{Tilted black hole accretion disc models of Sagittarius A*: time-variable millimetre to near-infrared emission}",
	journal = {\mnras},
	year = 2013,
	month = jul,
	volume = {432},
	number = {3},
	pages = {2252-2272},
	doi = {10.1093/mnras/stt583},
	archivePrefix = {arXiv},
	eprint = {1204.4454},
	primaryClass = {astro-ph.HE},
	adsurl = {https://ui.adsabs.harvard.edu/abs/2013MNRAS.432.2252D}
}

@ARTICLE{GRAVITY2020SgrAFlux,
	author = {{GRAVITY Collaboration}},
	title = "{The flux distribution of Sgr A*}",
	journal = {\aap},
	year = 2020,
	month = jun,
	volume = {638},
	eid = {A2},
	pages = {A2},
	doi = {10.1051/0004-6361/202037717},
	archivePrefix = {arXiv},
	eprint = {2004.07185},
	primaryClass = {astro-ph.GA},
	adsurl = {https://ui.adsabs.harvard.edu/abs/2020A&A...638A...2G}
}

@INPROCEEDINGS{Doeleman2019MoviesGround,
	author = {{Doeleman}, Sheperd and {Blackburn}, Lindy and {Dexter}, Jason and {Gomez}, Jose L. and {Johnson}, Michael D. and {Palumbo}, Daniel C. and {Weintroub}, Jonathan},
	title = "{Studying Black Holes on Horizon Scales with VLBI Ground Arrays}",
	booktitle = {\baas},
	year = 2019,
	volume = {51},
	month = sep,
	eid = {256},
	pages = {256},
	doi = {10.48550/arXiv.1909.01411},
	archivePrefix = {arXiv},
	eprint = {1909.01411},
	primaryClass = {astro-ph.IM},
	adsurl = {https://ui.adsabs.harvard.edu/abs/2019BAAS...51g.256D}
}

@INPROCEEDINGS{Johnson2019MoviesSpace,
	author = {{Johnson}, Michael and {Haworth}, Kari and {Pesce}, Dominic W. and {Palumbo}, Daniel C.~M. and {Blackburn}, Lindy},
	booktitle = {\baas},
	year = 2019,
	volume = {51},
	month = sep,
	eid = {235},
	pages = {235},
	doi = {10.48550/arXiv.1909.01405},
	archivePrefix = {arXiv},
	eprint = {1909.01405},
	primaryClass = {astro-ph.IM},
	adsurl = {https://ui.adsabs.harvard.edu/abs/2019BAAS...51g.235J}
}

@inproceedings{Johnson2024BHEX,
	title={The black hole explorer: Motivation and vision},
	author={Johnson, Michael D and Akiyama, Kazunori and Baturin, Rebecca and Bilyeu, Bryan and Blackburn, Lindy and Boroson, Don and C{\'a}rdenas-Avenda{\~n}o, Alejandro and Chael, Andrew and Chan, Chi-kwan and Chang, Dominic and others},
	booktitle={Space Telescopes and Instrumentation 2024: Optical, Infrared, and Millimeter Wave},
	volume={13092},
	pages={784--835},
	year={2024},
	organization={SPIE}
}

@ARTICLE{Johnson2023ngEHT,
	author = {{Johnson}, Michael D. and {Akiyama}, Kazunori and {Blackburn}, Lindy and {Bouman}, Katherine L. and {Broderick}, Avery E. and {Cardoso}, Vitor},
	title = "{Key Science Goals for the Next-Generation Event Horizon Telescope}",
	journal = {\galaxies},
	year = 2023,
	month = apr,
	volume = {11},
	number = {3},
	eid = {61},
	pages = {61},
	doi = {10.3390/galaxies11030061},
	archivePrefix = {arXiv},
	eprint = {2304.11188},
	primaryClass = {astro-ph.HE},
	adsurl = {https://ui.adsabs.harvard.edu/abs/2023Galax..11...61J}
}

@article{Nandra2013NewAthena,
title={The Hot and Energetic Universe: A White Paper presenting the science theme motivating the Athena+ mission},
author={Nandra, Kirpal and Barret, Didier and Barcons, Xavier and Fabian, Andy and Herder, Jan-Willem den and Piro, Luigi and Watson, Mike and Adami, Christophe and Aird, James and Afonso, Jose Manuel and others},
journal={arXiv preprint arXiv:1306.2307},
year={2013}
}

@ARTICLE{Porth2019GRMHD,
	author = {{Porth}, Oliver and {Chatterjee}, Koushik and {Narayan}, Ramesh and {Gammie}, Charles F. and {Mizuno}},
	title = "{The Event Horizon General Relativistic Magnetohydrodynamic Code Comparison Project}",
	journal = {\apjs},
	year = 2019,
	month = aug,
	volume = {243},
	number = {2},
	eid = {26},
	pages = {26},
	doi = {10.3847/1538-4365/ab29fd},
	archivePrefix = {arXiv},
	eprint = {1904.04923},
	primaryClass = {astro-ph.HE},
	adsurl = {https://ui.adsabs.harvard.edu/abs/2019ApJS..243...26P}
}

@ARTICLE{Mehlhaff2025PIC,
	author = {{Mehlhaff}, J. and {Cerutti}, B. and {Crinquand}, B.},
	title = "{A kinetic model of jet-corona coupling in accreting black holes}",
	journal = {\aap},
	year = 2025,
	month = sep,
	volume = {701},
	eid = {A62},
	pages = {A62},
	doi = {10.1051/0004-6361/202453561},
}

@ARTICLE{Nathanail2020Plasmoid,
	author = {{Nathanail}, Antonios and {Fromm}, Christian M. and {Porth}, Oliver and {Olivares}, Hector and {Younsi}, Ziri and {Mizuno}, Yosuke and {Rezzolla}, Luciano},
	title = "{Plasmoid formation in global GRMHD simulations and AGN flares}",
	journal = {\mnras},
	year = 2020,
	month = jun,
	volume = {495},
	number = {2},
	pages = {1549-1565},
	doi = {10.1093/mnras/staa1165},
	archivePrefix = {arXiv},
	eprint = {2002.01777},
	primaryClass = {astro-ph.HE},
	adsurl = {https://ui.adsabs.harvard.edu/abs/2020MNRAS.495.1549N}
}

@ARTICLE{Ripperda2020HotSpot,
	author = {{Ripperda}, Bart and {Bacchini}, Fabio and {Philippov}, Alexander A.},
	title = "{Magnetic Reconnection and Hot Spot Formation in Black Hole Accretion Disks}",
	journal = {\apj},
	year = 2020,
	month = sep,
	volume = {900},
	number = {2},
	eid = {100},
	pages = {100},
	doi = {10.3847/1538-4357/ababab},
}

@ARTICLE{Ficarra2026Inhomogeneities,
	author = {{Ficarra}, Giuseppe and {Arcuri}, Michele and {Megale}, Rita and {Servidio}, Sergio},
	title = "{The Role of Inhomogeneities in the Turbulent Accretion of Black Holes}",
	journal = {\apj},
	year = 2026,
	month = apr,
	volume = {1000},
	number = {2},
	eid = {186},
	pages = {186},
	doi = {10.3847/1538-4357/ae4d0e},
	archivePrefix = {arXiv},
	eprint = {2602.19104},
	primaryClass = {gr-qc},
	adsurl = {https://ui.adsabs.harvard.edu/abs/2026ApJ..1000..186F}
}

@ARTICLE{Ripperda2022Flare,
	author = {{Ripperda}, B. and {Liska}, M. and {Chatterjee}, K. and {Musoke}, G. and {Philippov}, A.~A. and {Markoff}, S.~B. and {Tchekhovskoy}, A. and {Younsi}, Z.},
	title = "{Black Hole Flares: Ejection of Accreted Magnetic Flux through 3D Plasmoid-mediated Reconnection}",
	journal = {\apjl},
	year = 2022,
	month = jan,
	volume = {924},
	number = {2},
	eid = {L32},
	pages = {L32},
	doi = {10.3847/2041-8213/ac46a1},
	archivePrefix = {arXiv},
	eprint = {2109.15115},
	primaryClass = {astro-ph.HE},
	adsurl = {https://ui.adsabs.harvard.edu/abs/2022ApJ...924L..32R}
}

@ARTICLE{Lee2021Inhomogeneities,
	author = {{Lee}, Daeyoung and {Gammie}, Charles F.},
	title = "{Disks as Inhomogeneous, Anisotropic Gaussian Random Fields}",
	journal = {\apj},
	year = 2021,
	month = jan,
	volume = {906},
	number = {1},
	eid = {39},
	pages = {39},
	doi = {10.3847/1538-4357/abc8f3},
	archivePrefix = {arXiv},
	eprint = {2011.07151},
	primaryClass = {astro-ph.IM},
	adsurl = {https://ui.adsabs.harvard.edu/abs/2021ApJ...906...39L}
}

@ARTICLE{Chesler2021Autocorrelations,
	author = {{Chesler}, Paul M. and {Blackburn}, Lindy and {Doeleman}, Sheperd S. and {Johnson}, Michael D. and {Moran}, James M. and {Narayan}, Ramesh and {Wielgus}, Maciek},
	title = "{Light echos and coherent autocorrelations in a black hole spacetime}",
	journal = {\cqg},
	year = 2021,
	month = jun,
	volume = {38},
	number = {12},
	eid = {125006},
	pages = {125006},
	doi = {10.1088/1361-6382/abeae4},
	archivePrefix = {arXiv},
	eprint = {2012.11778},
	primaryClass = {gr-qc},
	adsurl = {https://ui.adsabs.harvard.edu/abs/2021CQGra..38l5006C}
}

@ARTICLE{Wong2021Glimmer,
	author = {{Wong}, George N.},
	title = "{Black Hole Glimmer Signatures of Mass, Spin, and Inclination}",
	journal = {\apj},
	year = 2021,
	month = mar,
	volume = {909},
	number = {2},
	eid = {217},
	pages = {217},
	doi = {10.3847/1538-4357/abdd2d},
	archivePrefix = {arXiv},
	eprint = {2009.06641},
	primaryClass = {astro-ph.HE},
	adsurl = {https://ui.adsabs.harvard.edu/abs/2021ApJ...909..217W}
}

@ARTICLE{Bezdvekovz2026Autocorrelations,
	author = {{Bezd{\v{e}}kov{\'a}}, Barbora and {Hadar}, Shahar and {Wong}, George N. and {Wielgus}, Maciek},
	title = "{Extreme lensing signatures revealed by correlations of simulated black hole movies}",
	journal = {Nature Astronomy},
	year = 2026,
	month = may,
	doi = {10.1038/s41550-026-02874-x},
}

@ARTICLE{Gralla2020Lensing,
	author = {{Gralla}, Samuel E. and {Lupsasca}, Alexandru},
	title = "{Lensing by Kerr black holes}",
	journal = {\prd},
	year = 2020,
	month = feb,
	volume = {101},
	number = {4},
	eid = {044031},
	pages = {044031},
	doi = {10.1103/PhysRevD.101.044031},
	archivePrefix = {arXiv},
	eprint = {1910.12873},
	primaryClass = {gr-qc},
	adsurl = {https://ui.adsabs.harvard.edu/abs/2020PhRvD.101d4031G}
}

@ARTICLE{Hadar2021PhotonRingAutocorrelation,
	author = {{Hadar}, Shahar and {Johnson}, Michael D. and {Lupsasca}, Alexandru and {Wong}, George N.},
	title = "{Photon ring autocorrelations}",
	journal = {\prd},
	year = 2021,
	month = may,
	volume = {103},
	number = {10},
	eid = {104038},
	pages = {104038},
	doi = {10.1103/PhysRevD.103.104038},
	archivePrefix = {arXiv},
	eprint = {2010.03683},
	primaryClass = {gr-qc},
	adsurl = {https://ui.adsabs.harvard.edu/abs/2021PhRvD.103j4038H}
}

@ARTICLE{Wong2024EchoesVLBI,
	author = {{Wong}, George N. and {Medeiros}, Lia and {C{\'a}rdenas-Avenda{\~n}o}, Alejandro and {Stone}, James M.},
	title = "{Measuring Black Hole Light Echoes with Very Long Baseline Interferometry}",
	journal = {\apjl},
	year = 2024,
	month = nov,
	volume = {975},
	number = {2},
	eid = {L40},
	pages = {L40},
	doi = {10.3847/2041-8213/ad8650},
	archivePrefix = {arXiv},
	eprint = {2410.10950},
	primaryClass = {astro-ph.HE},
	adsurl = {https://ui.adsabs.harvard.edu/abs/2024ApJ...975L..40W}
}

@ARTICLE{Cardenas2024Autocorrelations,
	author = {{C{\'a}rdenas-Avenda{\~n}o}, Alejandro and {Gammie}, Charles and {Lupsasca}, Alexandru},
	title = "{Explanation for the Absence of Secondary Peaks in Black Hole Light Curve Autocorrelations}",
	journal = {\prl},
	year = 2024,
	month = sep,
	volume = {133},
	number = {13},
	eid = {131402},
	pages = {131402},
	doi = {10.1103/PhysRevLett.133.131402},
	archivePrefix = {arXiv},
	eprint = {2406.04176},
	primaryClass = {astro-ph.HE},
	adsurl = {https://ui.adsabs.harvard.edu/abs/2024PhRvL.133m1402C}
}

@ARTICLE{Hadar2023Spectrotemporal,
	author = {{Hadar}, Shahar and {Harikesh}, Sreehari and {Chelouche}, Doron},
	title = "{Extreme lensing induces spectrotemporal correlations in black-hole signals}",
	journal = {\prd},
	year = 2023,
	month = jun,
	volume = {107},
	number = {12},
	eid = {124057},
	pages = {124057},
	doi = {10.1103/PhysRevD.107.124057},
	archivePrefix = {arXiv},
	eprint = {2305.11247},
	primaryClass = {gr-qc},
	adsurl = {https://ui.adsabs.harvard.edu/abs/2023PhRvD.107l4057H}
}

@ARTICLE{Harikesh2025Spectrotemporal,
	author = {{Harikesh}, Sreehari and {Hadar}, Shahar and {Chelouche}, Doron},
	title = "{Exploring lensing signatures through spectrotemporal correlations: Implications for black hole parameter estimation}",
	journal = {\prd},
	year = 2025,
	month = aug,
	volume = {112},
	number = {4},
	eid = {043020},
	pages = {043020},
	doi = {10.1103/wjpm-9byt},
	archivePrefix = {arXiv},
	eprint = {2502.12053},
	primaryClass = {astro-ph.HE},
	adsurl = {https://ui.adsabs.harvard.edu/abs/2025PhRvD.112d3020H}
}

@ARTICLE{Vincent2011gyoto,
	author = {{Vincent}, F.~H. and {Paumard}, T. and {Gourgoulhon}, E. and {Perrin}, G.},
	title = "{GYOTO: a new general relativistic ray-tracing code}",
	journal = {\cqg},
	year = 2011,
	month = nov,
	volume = {28},
	number = {22},
	eid = {225011},
	pages = {225011},
	doi = {10.1088/0264-9381/28/22/225011},
	archivePrefix = {arXiv},
	eprint = {1109.4769},
	primaryClass = {gr-qc},
	adsurl = {https://ui.adsabs.harvard.edu/abs/2011CQGra..28v5011V}
}

@ARTICLE{GRAVITY2021Distance,
	author = {{GRAVITY Collaboration} and {Abuter}, R. and {Amorim}, A. and {Baub{\"o}ck}, M. and {Berger}, J.~P. and {Bonnet}, H. and {Brandner}, W. and {Cl{\'e}net}, Y.},
	title = "{Improved GRAVITY astrometric accuracy from modeling optical aberrations}",
	journal = {\aap},
	year = 2021,
	month = mar,
	volume = {647},
	eid = {A59},
	pages = {A59},
	doi = {10.1051/0004-6361/202040208},
	archivePrefix = {arXiv},
	eprint = {2101.12098},
	primaryClass = {astro-ph.GA},
	adsurl = {https://ui.adsabs.harvard.edu/abs/2021A&A...647A..59G}
}

@ARTICLE{GRAVITY2022Mass,
	author = {{GRAVITY Collaboration} and {Abuter}, R. and {Aimar}, N. and {Amorim}, A. and {Ball}, J. and {Baub{\"o}ck}, M. and {Berger}, J.~P. and {Bonnet}, H. and {Bourdarot}, G. and {Brandner}, W. and {Cardoso}, V. and {Cl{\'e}net}, Y.},
	title = "{Mass distribution in the Galactic Center based on interferometric astrometry of multiple stellar orbits}",
	journal = {\aap},
	year = 2022,
	month = jan,
	volume = {657},
	eid = {L12},
	pages = {L12},
	doi = {10.1051/0004-6361/202142465},
	archivePrefix = {arXiv},
	eprint = {2112.07478},
	primaryClass = {astro-ph.GA},
	adsurl = {https://ui.adsabs.harvard.edu/abs/2022A&A...657L..12G}
}

@ARTICLE{GRAVITY2018Inclination,
	author = {{GRAVITY Collaboration} and {Abuter}, R. and {Amorim}, A. and {Baub{\"o}ck}, M. and {Berger}, J.~P. and {Bonnet}, H. and {Brandner}, W. and {Cl{\'e}net}, Y.},
	title = "{Detection of orbital motions near the last stable circular orbit of the massive black hole SgrA*}",
	journal = {\aap},
	year = 2018,
	month = oct,
	volume = {618},
	eid = {L10},
	pages = {L10},
	doi = {10.1051/0004-6361/201834294},
	archivePrefix = {arXiv},
	eprint = {1810.12641},
	primaryClass = {astro-ph.GA},
	adsurl = {https://ui.adsabs.harvard.edu/abs/2018A&A...618L..10G}
}

@ARTICLE{Aimar2023Flares,
       author = {{Aimar}, N. and {Dmytriiev}, A. and {Vincent}, F.~H. and {El Mellah}, I. and {Paumard}, T. and {Perrin}, G. and {Zech}, A.},
        title = "{Magnetic reconnection plasmoid model for Sagittarius A* flares}",
      journal = {\aap},
         year = 2023,
        month = apr,
       volume = {672},
          eid = {A62},
        pages = {A62},
          doi = {10.1051/0004-6361/202244936},
archivePrefix = {arXiv},
       eprint = {2301.11874},
 primaryClass = {astro-ph.HE},
       adsurl = {https://ui.adsabs.harvard.edu/abs/2023A&A...672A..62A}
}

@ARTICLE{Bardeen1972Kerr,
	author = {{Bardeen}, James M. and {Press}, William H. and {Teukolsky}, Saul A.},
	title = "{Rotating Black Holes: Locally Nonrotating Frames, Energy Extraction, and Scalar Synchrotron Radiation}",
	journal = {\apj},
	year = 1972,
	month = dec,
	volume = {178},
	pages = {347-370},
	doi = {10.1086/151796},
	adsurl = {https://ui.adsabs.harvard.edu/abs/1972ApJ...178..347B}
}

@ARTICLE{VanderPlas2018LombScargle,
	author = {{VanderPlas}, Jacob T.},
	title = "{Understanding the Lomb-Scargle Periodogram}",
	journal = {\apjs},
	year = 2018,
	month = may,
	volume = {236},
	number = {1},
	eid = {16},
	pages = {16},
	doi = {10.3847/1538-4365/aab766},
	archivePrefix = {arXiv},
	eprint = {1703.09824},
	primaryClass = {astro-ph.IM},
	adsurl = {https://ui.adsabs.harvard.edu/abs/2018ApJS..236...16V}
}

@BOOK{MTW1973Gravitation,
	author = {{Misner}, Charles W. and {Thorne}, Kip S. and {Wheeler}, John Archibald},
	title = "{Gravitation}",
	year = 1973,
	adsurl = {https://ui.adsabs.harvard.edu/abs/1973grav.book.....M},
	publisher = {Freeman San Francisco}
}

@book{Rybicki&Lightman1979Radiative,
	title={Radiative processes in astrophysics},
	author={Rybicki, George B and Lightman, Alan P},
	year={1979},
	publisher={John Wiley \& Sons}
}

@article{Lindquist1966Transfer,
	title={Relativistic transport theory},
	author={Lindquist, Richard W},
	journal={\annphys},
	volume={37},
	number={3},
	pages={487--518},
	year={1966},
	publisher={Elsevier}
}

@book{bannikova2022foundations,
	title={Foundations of celestial mechanics},
	author={Bannikova, Elena and Capaccioli, Massimo},
	year={2022},
	publisher={Springer}
}

@ARTICLE{Urso2025PhotonRings,
	author = {{Urso}, I. and {Vincent}, F.~H. and {Wielgus}, M. and {Paumard}, T. and {Perrin}, G.},
	title = "{Gravity versus astrophysics in black hole images and photon rings: Equatorial emissions and spherically symmetric space-times}",
	journal = {\aap},
	year = 2025,
	month = aug,
	volume = {700},
	eid = {A193},
	pages = {A193},
	doi = {10.1051/0004-6361/202554919},
}

@ARTICLE{Chowdhury2026TimeDelayAsymmetry,
	author = {{Chowdhury}, Shakibul},
	title = "{A New Timing Signature of Black Hole Spin: Time-Delay Asymmetry in Kerr Accretion Flows}",
	journal = {arXiv e-prints},
	year = 2026,
	month = may,
	eid = {arXiv:2605.11734},
	pages = {arXiv:2605.11734},
	doi = {10.48550/arXiv.2605.11734},
}

@ARTICLE{RojasPaternina2026LightPropagation,
	author = {{Rojas-Paternina}, Daniel and {C{\'a}rdenas-Avenda{\~n}o}, Alejandro},
	title = "{Light Propagation Prescriptions for Black Hole Movies}",
	journal = {arXiv e-prints},
	year = 2026,
	month = may,
	eid = {arXiv:2605.12659},
	pages = {arXiv:2605.12659},
	doi = {10.48550/arXiv.2605.12659},
}

@ARTICLE{Cardenas2023AART,
       author = {{C{\'a}rdenas-Avenda{\~n}o}, Alejandro and {Lupsasca}, Alexandru and {Zhu}, Hengrui},
        title = "{Adaptive analytical ray tracing of black hole photon rings}",
      journal = {\prd},
         year = 2023,
        month = feb,
       volume = {107},
       number = {4},
          eid = {043030},
        pages = {043030},
          doi = {10.1103/PhysRevD.107.043030},
archivePrefix = {arXiv},
       eprint = {2211.07469},
 primaryClass = {gr-qc},
       adsurl = {https://ui.adsabs.harvard.edu/abs/2023PhRvD.107d3030C}
}

@article{Lampa1924SR,
	title={Wie erscheint nach der Relativit{\"a}tstheorie ein bewegter Stab einem ruhenden Beobachter? (According to the theory of relativity, how does a moving rod appear to a stationary observer?)},
	author={Lampa, Anton},
	journal={Zeitschrift f{\"u}r Physik},
	volume={27},
	number={1},
	pages={138--148},
	year={1924},
	publisher={Springer}
}

@article{Penrose1959SR,
	title={The apparent shape of a relativistically moving sphere},
	author={Penrose, Roger},
	journal={Mathematical proceedings of the Cambridge philosophical society},
	volume={55},
	number={1},
	pages={137--139},
	year={1959},
	organization={Cambridge University Press}
}

@article{Terrell1959SR,
	title={Invisibility of the Lorentz contraction},
	author={Terrell, James},
	journal={Physical Review},
	volume={116},
	number={4},
	pages={1041},
	year={1959},
	publisher={APS}
}

@ARTICLE{Mummery2026TDT,
	author = {{Mummery}, Andrew and {Guo}, Minghao and {Quataert}, Eliot and {Stone}, James M.},
	title = "{The long-term evolution of a black hole accretion torus: a comparison between GRMHD simulations and relativistic thin disk theory}",
	journal = {arXiv e-prints},
	year = 2026,
	month = oct,
	eid = {arXiv:2610.35648},
	pages = {arXiv:2610.35648},
	doi = {10.48550/arXiv.2610.35648},
	archivePrefix = {arXiv},
	eprint = {2610.35648},
	primaryClass = {astro-ph.HE},
	adsurl = {https://ui.adsabs.harvard.edu/abs/2026arXiv261035648M}
}
	
	\begin{appendix} 
	
		\section{Emission coordinates}
		\label{app:hit}
\begin{figure*}[ht!]
			\centering
			\resizebox{18cm}{!}
			{\includegraphics{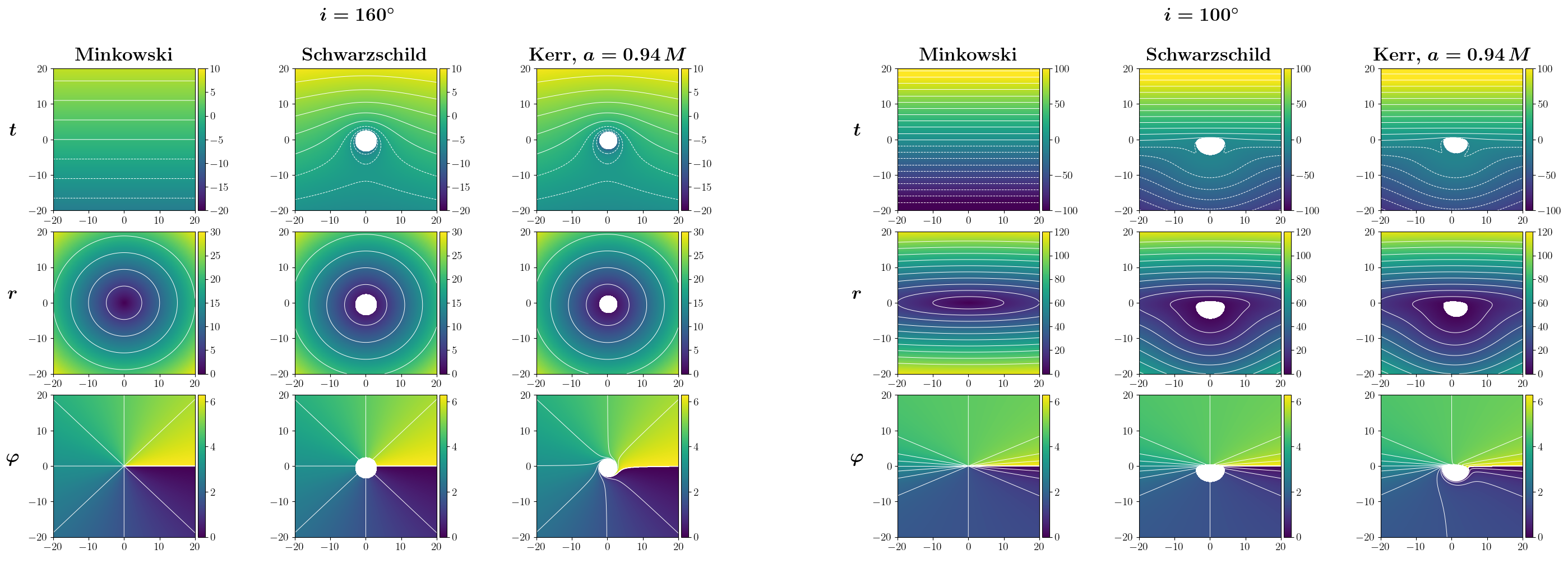}}
			\caption{Impact-coordinate maps for two viewing inclinations: $i=160^\circ$ (left panel) and $i=100^\circ$ (right panel). In each panel, the rows correspond to the time ($t$, top), radial ($r$, middle), and azimuthal ($\varphi$, bottom) coordinates, while the columns show results for Minkowski (left), Schwarzschild (centre), and Kerr with $a=0.94$ (right). White lines mark contours at fixed coordinate value (solid for positive values,  dashed for negative values). For $i=160^\circ$, contours are spaced by $\Delta t=2\,M$ over the range $-6\,M \le t \le 8\,M$, $\Delta r=5\,M$ over $0\,M \le r \le 25\,M$, and $\Delta\varphi=\pi/4$ over $0 \le \varphi < 2\pi$. For $i=100^\circ$, contours are spaced by $\Delta t=10\,M$ over $-100\,M \le t \le 110\,M$, $\Delta r=10\,M$ over $0\,M \le r \le 100\,M$, and $\Delta\varphi=\pi/8$ over $0 \le \varphi < 2\pi$.}
			\label{fig:coordmetric}
		\end{figure*}

		We present in Figure \hyperref[fig:coordmetric]{\ref{fig:coordmetric}} 
		the full-screen impact coordinates for the two viewing inclinations $i=160^\circ$ and $i=100^\circ$, and for a screen's northern direction aligned with the projected angular-momentum vector of the source. In Minkowski space-time, the maps exhibit both top-bottom and left-right reflection symmetries. In the Schwarzschild case, only the left-right symmetry is preserved, while in Kerr space-time no such symmetry remains. In curved space-times, the emission times increase towards the black hole and on the side of the disc further from the observer, as photons must undergo stronger deflection to escape to infinity. In the azimuthal coordinate for the Kerr case, the asymmetry reflects the Lense-Thirring effect, that is, the dragging of space-time by the black hole rotation. We note that the properties of Kerr time maps have attracted considerable interest in recent studies \citep{Chowdhury2026TimeDelayAsymmetry,RojasPaternina2026LightPropagation}.
		
		\section{Minkowski analytical derivations} 
		\label{app:Minko}
		\subsection{Rømer delay}
		We derive equation \eqref{eq:FourierSeries}, following the steps used in the solution finding of Kepler's equation, $M=E-e\sin E$, which is transcendental in $E$, as explained in Appendix F of \citep{bannikova2022foundations}. We start from equation \eqref{eq:flight} that we retranscript for convenience:
		\begin{subequations}
		\begin{align}
			t^\mathrm{\,em,\,c}_\mathrm{Minko}&=\tilde{t}^\mathrm{\:obs,\,c}_\mathrm{Minko}-r_\mathrm{c}\sin\iota\sin(\Omega^\mathrm{c}_\mathrm{orb}t^\mathrm{\,em,\,c}_\mathrm{Minko}+\varphi^\mathrm{\,c}_0)\\
			\Rightarrow y&=x+\alpha f(y)\:,\label{eq:Lagrange}
		\end{align} 
	\end{subequations}
		where we have identified
	\begingroup
	\setcounter{equation}{\value{equation}-1}
	\begin{subequations}
		\renewcommand{\theequation}{\theparentequation\alph{equation}}
		\setcounter{equation}{2} 
		\begin{align}
			y&=\Omega^\mathrm{c}_\mathrm{orb}t^\mathrm{\,em,\,c}_\mathrm{Minko}+\varphi^\mathrm{\,c}_0\\
			x&=\Omega^\mathrm{c}_\mathrm{orb}\tilde{t}^\mathrm{\:obs,\,c}_\mathrm{Minko}+\varphi^\mathrm{\,c}_0\\
			\alpha&=-\Omega^\mathrm{c}_\mathrm{orb}r_\mathrm{c}\sin\iota\\
			f&:y\mapsto \sin{y}\:.
		\end{align}
		\end{subequations}
	\endgroup
		\indent Lagrange inversion theorem states that any infinitely derivable function $g:y\mapsto g(y)$, whose independent variable $y$ satisfies a relation of the type \eqref{eq:Lagrange}, with $f$ infinitely derivable, can be expanded in a power series of $\alpha$:
		\begin{align}
			g(y)=g(x)+\sum_{n=1}^\infty\frac{\alpha^n}{n!}\frac{\mathrm{d}^{n-1}}{\mathrm{d}x^{n-1}}\left(f^n(x)\frac{\mathrm{d}g(x)}{\mathrm{d}x}\right)\:.
		\end{align}
		Since our final goal is to find $t^\mathrm{\,em,\,c}_\mathrm{Minko}$, which appears linearly in $y$, we need to expand for the identity $g:y\mapsto y$. Thus, in our case, Lagrange inversion theorem leads to
		\begin{align} y-x&=\sum_{n=1}^\infty\frac{\alpha^n}{n!}\frac{\mathrm{d}^{n-1}}{\mathrm{d}x^{n-1}}\left(\sin^n(x)\,\right)\:.\label{eq:odd}
		\end{align}
		The left term, $y-x$, is then a $2\pi$-periodic odd function of $x$.\footnote{As it can be proven easily by induction, every term of the sum is odd.}\\
		\indent The Fourier theorem asserts that if $h(x)$ is a odd periodic function of period $P$, it can be expanded in a trigonometric series:
		\begin{subequations}
		\begin{align}
			h(x)&=\sum_{n=1}^\infty b_n\sin\left(2\pi\frac{n}{P}x\right)\:,\\ b_n&=\frac{4}{P}\int_0^{P/2} h(x)\sin\left(2\pi\frac{n}{P}x\right)\mathrm{d}x\:.
		\end{align}
		\end{subequations}
		Equation \eqref{eq:odd} can then be rewritten as
		\begin{subequations}
		\begin{align}
			y-x&=\sum_{n=1}^\infty b_n\sin\left(nx\right)\:,\\ b_n&=\frac{2}{\pi}\int_0^{\pi} (y-x)\sin\left(nx\right)\mathrm{d}x\:,
			\label{eq:oddsum}
		\end{align}
		\end{subequations}
	     where the coefficients $b_n$ can be integrated by parts:
		\begin{align}
			b_n &=\frac{2}{n\pi}\int_0^{\pi}\cos\left(ny-n\alpha\sin y\right)\,\mathrm{d}y\:.
			\label{eq:bn}
		\end{align}
		The $n$-th Bessel function of the first kind is defined as
		\begin{align}
			J_n(x)\equiv\frac{1}{\pi}\int_0^{\pi}\cos\left(nt-x\sin t\right)\,\mathrm{d}t\:.
		\end{align}
		It is then straightforward to rewrite \eqref{eq:bn} as
		\begin{align}
			b_n=\frac{2}{n}J_n(n\alpha)\:.\label{eq:Bessel}
		\end{align}
		\indent Thus, combining equations \eqref{eq:oddsum} and \eqref{eq:Bessel}, we finally get:
		\begin{align}
			y&=x+\sum_{n=1}^\infty\frac{2}{n}J_n(n\alpha)\sin(nx)
		\end{align}
		and reintroducing the full expressions and using the fact that $J_n(-x)=(-1)^nJ_n(x)$, one retrieves equation \eqref{eq:FourierSeries}.
		
		\subsection{Redshift}
		In Minkowski space-time, the redshift factor for usual velocity profiles can be easily expressed analytically \citep{Urso2025PhotonRings} since the 4-momentum of the photon, $\mathbf{p}\equiv\mathbf{p}^\mathrm{obs}=\mathbf{p}^\mathrm{em}$, has a simple, geodesic invariant, analytical form:  
		\begin{align}
			\mathbf{p}=\mathbf{e}_t-\sin\iota\sin\varphi^\mathrm{em}\,\mathbf{e}_r+\cos\iota\,\mathbf{e}_\theta-\sin\iota\cos\varphi^\mathrm{em}\, \mathbf{e}_\varphi\:,\label{eq:pem}
		\end{align}
		where $\iota=\pi-i\in[0^\circ;90^\circ]$ for $i\in[90^\circ;180^\circ]$ and where the natural basis vectors, $\boldsymbol{\partial}_\mu$, have been replaced by
		\begin{align}
			\mathbf{e}_t=\frac{\boldsymbol{\partial}_t}{\sqrt{-g_{tt}}},\quad\mathbf{e}_r=\frac{\boldsymbol{\partial}_r}{\sqrt{g_{rr}}},\quad\mathbf{e}_\theta=\frac{\boldsymbol{\partial}_\theta}{\sqrt{g_{\theta\theta}}},\quad\mathbf{e}_\varphi=\frac{\boldsymbol{\partial}_\varphi}{\sqrt{g_{\varphi\varphi}}}\:.
		\end{align}
		\indent As the velocity of a static observer is $\mathbf{u}^\mathrm{obs}=\mathbf{e}_t$, one has $\mathbf{p}^\mathrm{obs}\cdot \mathbf{u}^\mathrm{obs}=-1$ at the numerator of equation \eqref{eq:redshift}.
		The general expressions of the hot spot's circular 4-velocity is
		\begin{align}
			\mathbf{u}^\mathrm{em}&=u^{t,\,\mathrm{c}}_\mathrm{em}(\mathbf{\partial}_t+\Omega^\mathrm{c}_\mathrm{orb}\mathbf{\partial}_\varphi)\\\Omega^\mathrm{c}_\mathrm{orb}&=u^{\varphi,\,\mathrm{c}}_\mathrm{em}/u^{t,\,\mathrm{c}}_\mathrm{em}=\mathrm{d}\varphi^\mathrm{\,c}_\mathrm{em}/\mathrm{d}t^\mathrm{\,c}_\mathrm{em}\:.
		\end{align}
		The expression of $\Omega^\mathrm{c}_\mathrm{orb}$ is provided by equation \eqref{eq:Omega} and $u^{t,\,\mathrm{c}}_\mathrm{em}$ can be deduced by the normalisation $\mathbf{u}^\mathrm{em}\cdot\mathbf{u}^\mathrm{em}=-1$:
		\begin{align}
			\Omega^\mathrm{c}_\mathrm{orb}=\frac{\sqrt{M}}{r_\mathrm{c}^{3/2}}\quad\Rightarrow\quad u^{t,\,\mathrm{c}}_\mathrm{em}=\sqrt{\frac{r_\mathrm{c}}{r_\mathrm{c}-M}}\:.
		\end{align}
		Thus, in the orthonormal basis $\mathbf{e}_\mu$, the emitter 4-velocity reads
		\begin{align}
			\mathbf{u}^\mathrm{em}=\sqrt{\frac{r_\mathrm{c}}{r_\mathrm{c}-M}}\left(\mathbf{e}_t+\sqrt{\frac{M}{r_\mathrm{c}}}\mathbf{e}_\varphi\right)\:.\label{eq:uem}
		\end{align}
		\indent With \eqref{eq:pem} and \eqref{eq:uem}, it is immediate to obtain the expressions of the denominator of $g$:
		\begin{align}
			\mathbf{p}^\mathrm{em}\cdot \mathbf{u}^\mathrm{em}=-\sqrt{\frac{r_\mathrm{c}}{r_\mathrm{c}-M}}\left(1+\sqrt{\frac{M}{r_\mathrm{c}}}\sin\iota\cos\varphi^\mathrm{\,c}_\mathrm{em} \right)
		\end{align}
		and the analytical formula of the redshift is therefore given by equation \eqref{eq:redshiftMinko}.
		
		\section{Phase-modulated signals}
		
		\label{app:PM}
		In this appendix, we derive the decomposition of the phase-modulated sinusoids appearing in \eqref{eq:multiPM} and \eqref{eq:redshiftMinko}, that we denote $S^\mathrm{em}$ and $S^\mathrm{g}$, into a sum of harmonic components. We first consider $S^\mathrm{em}$ and discuss the straightforward modifications required for $S^\mathrm{g}$ afterwards. For readability, we define $\tilde{t}\equiv\tilde{t}^\mathrm{\:obs,\,c}_\mathrm{Minko}$. We begin by rewriting the sinusoid using complex exponentials via Euler’s formula and standard properties:
		\begin{align}
			S^\mathrm{em}&=\sin(2\pi f_\mathrm{mod}\tilde{t}+\varphi^\mathrm{\,c}_0+\sum^\infty_{n=1}\Phi_n\sin[2n\pi f_\mathrm{orb}^\mathrm{\,c}\tilde{t}+n\varphi^\mathrm{\,c}_0])\\
			&=\operatorname{Im}\left\{ e^{i2\pi f_\mathrm{mod}\tilde{t}+i\varphi^\mathrm{\,c}_0}\prod_{n=1}^\infty e^{i\,\Phi_n\sin[2n\pi f_\mathrm{orb}^\mathrm{\,c}\tilde{t}+n\varphi^\mathrm{\,c}_0]}\right\}\:.
		\end{align}
		We then apply Jacobi-Anger identity, $e^{iz\sin\theta}=\sum_{k=-\infty}^\infty J_k(z)e^{ik\theta}$:
		\begin{align}
			S^\mathrm{em}&=\operatorname{Im}\left\{ e^{i2\pi f_\mathrm{mod}\tilde{t}+i\varphi^\mathrm{\,c}_0}\prod_{n=1}^\infty \left[ \sum_{k_n=-\infty}^\infty J_{k_n}(\Phi_n)e^{i\,k_n[2n\pi f_\mathrm{orb}^\mathrm{c}\tilde{t}+n\varphi^\mathrm{\,c}_0]}\right] \right\}
		\end{align}
		and, expanding the product of sums, we obtain
		\begin{align}
		S^\mathrm{em}&=\sum_{k_1, k_2, \dots = -\infty}^{\infty} \left[ \prod_{n=1}^{\infty} J_{k_n}(\Phi_n) \right]\operatorname{Im}\left\{ e^{i2\pi f_\mathrm{mod}\tilde{t}+i\varphi^\mathrm{\,c}_0+i\sum_{n=1}^\infty nk_n[2\pi f_\mathrm{orb}^\mathrm{\,c}\tilde{t}+\varphi^\mathrm{\,c}_0]}\right\}\\
		&=\sum_{\boldsymbol{k}\in\mathbb{Z}^\infty} A_{\boldsymbol{k}}\sin(2\pi[ f_\mathrm{mod}+m_{\boldsymbol{k}} f_\mathrm{orb}^\mathrm{\,c}]\,\tilde{t}+[m_{\boldsymbol{k}}+1]\varphi^\mathrm{\,c}_0])\:,
	\end{align}
	where in the second equality we have defined:
	\begin{align}
		\boldsymbol{k}\equiv(k_1, k_2, \dots)\:,\quad A_{\boldsymbol{k}}\equiv \prod_{n=1}^{\infty} J_{k_n}(\Phi_n)\:,\quad m_{\boldsymbol{k}}\equiv\sum_{n=1}^\infty nk_n\:.
	\end{align}
	For the redshift signal, the carrier and modulation frequencies coincide, so the same frequency enters both roles. The derivation proceeds analogously, except that the real part is taken instead of the imaginary part, yielding:
	\begin{align}
		S^g&=\cos\left(2\pi f_\mathrm{orb}^\mathrm{\,c}\tilde{t}+\varphi_0^\mathrm{\,c}+\sum^\infty_{n=1}\bar{\Phi}_n\sin[2n\pi f_\mathrm{orb}^\mathrm{\,c}\tilde{t}+n\varphi^\mathrm{\,c}_0]\right)\:,\\
		&=\sum_{\boldsymbol{k}\in\mathbb{Z}^\infty} \bar{A}_{\boldsymbol{k}}\cos(2\pi[m_{\boldsymbol{k}}+1] f_\mathrm{orb}^\mathrm{\,c}\,\tilde{t}+[m_{\boldsymbol{k}}+1]\varphi^\mathrm{\,c}_0)\:,
	\end{align}
	where we have introduced
		\begin{align}
		\bar{A}_{\boldsymbol{k}}\equiv \prod_{n=1}^{\infty} J_{k_n}(\bar{\Phi}_n)\:.
	\end{align}
	
	\section{Supplementary material}
	\label{app:additional}
    We present here additional plots of the quantities of interest considered in the main text, either to visualise the impacts of varying model parameters or to isolate specific effects and support the interpretation of the spectral behaviour detailed in Section \hyperref[sec:4]{\ref{sec:4}}.
    
    \subsection{Areal size evolution}
    \label{sec:area}
    
    \begin{figure*}[ht!]
    	\centering
    	\resizebox{15cm}{!}
    	{\includegraphics{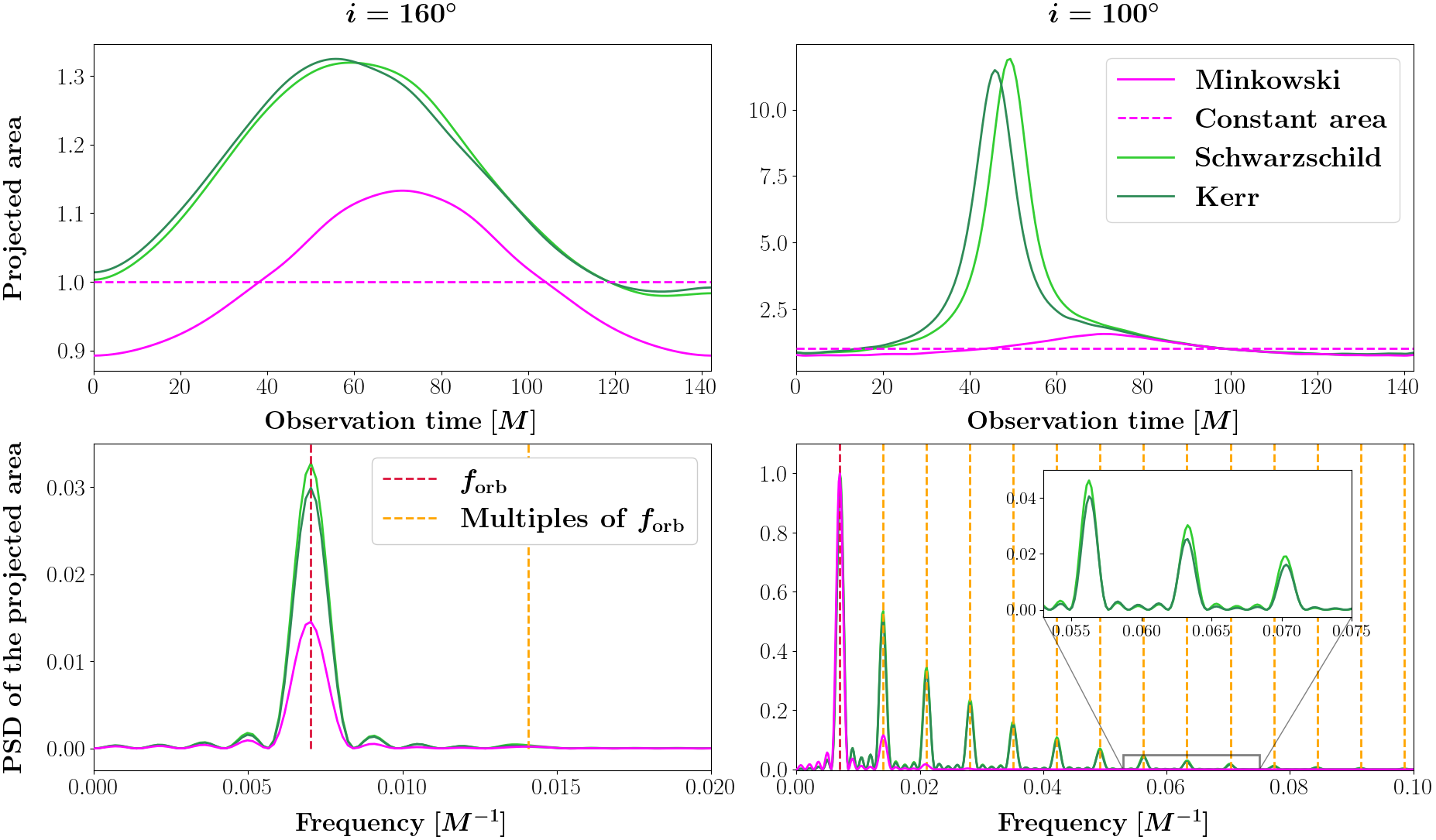}}
    	\caption{Hot spot’s areal size on the observer’s screen at $i = 160^\circ$ (top left panel) and $i = 100^\circ$  (top right panel) in Minkowski (magenta), Schwarzschild (light green), and Kerr ($a = 0.94$ $M$, dark green) space-times, alongside their associated PSDs (bottom panels). The areal sizes are normalised by the constant projected area due solely to the inclination in
    		flat space-time, shown as the horizontal dashed magenta line, and are smoothed with a Gaussian
    		kernel to suppress pixelisation-induced discontinuities. In the PSDs, the orbital frequency (red)
    		and its harmonics (orange) are indicated by dashed vertical lines. For visualisation, the PSDs at $i = 100^\circ$ are normalised to the Minkowski maximum for Minkowski space-time and to the Schwarzschild maximum for both Schwarzschild and Kerr space-times. Compared to Minkowski space-time, the areal projection is more asymmetric in curved space-time, producing more harmonics with greater relative power.}
    	\label{fig:terrell}
    \end{figure*}
    
    The variation in the source's apparent size is an intrinsic feature of ray-tracing simulations and cannot be switched off. However, its contribution can be isolated by assuming a constant emissivity and neglecting redshift effects, as done in Figure \hyperref[fig:terrell]{\ref{fig:terrell}}. We note that, although the functional behaviour of the Minkowski distortion already differs significantly from that in curved space-times at moderate inclinations, the difference in spectral behaviour becomes appreciable only at high inclinations. At these viewing angles, the lack of symmetry in the areal projection over the orbital period produces enhanced power in the harmonic channels. As highlighted in Section \hyperref[sec:4]{\ref{sec:4}}, this results in broader PSD envelopes and eventually produces interference between the orbital and source modulation combs.
	
	\subsection{Role of inclination}
	
\begin{figure}[h!]
		\centering
		\includegraphics[width=\hsize]{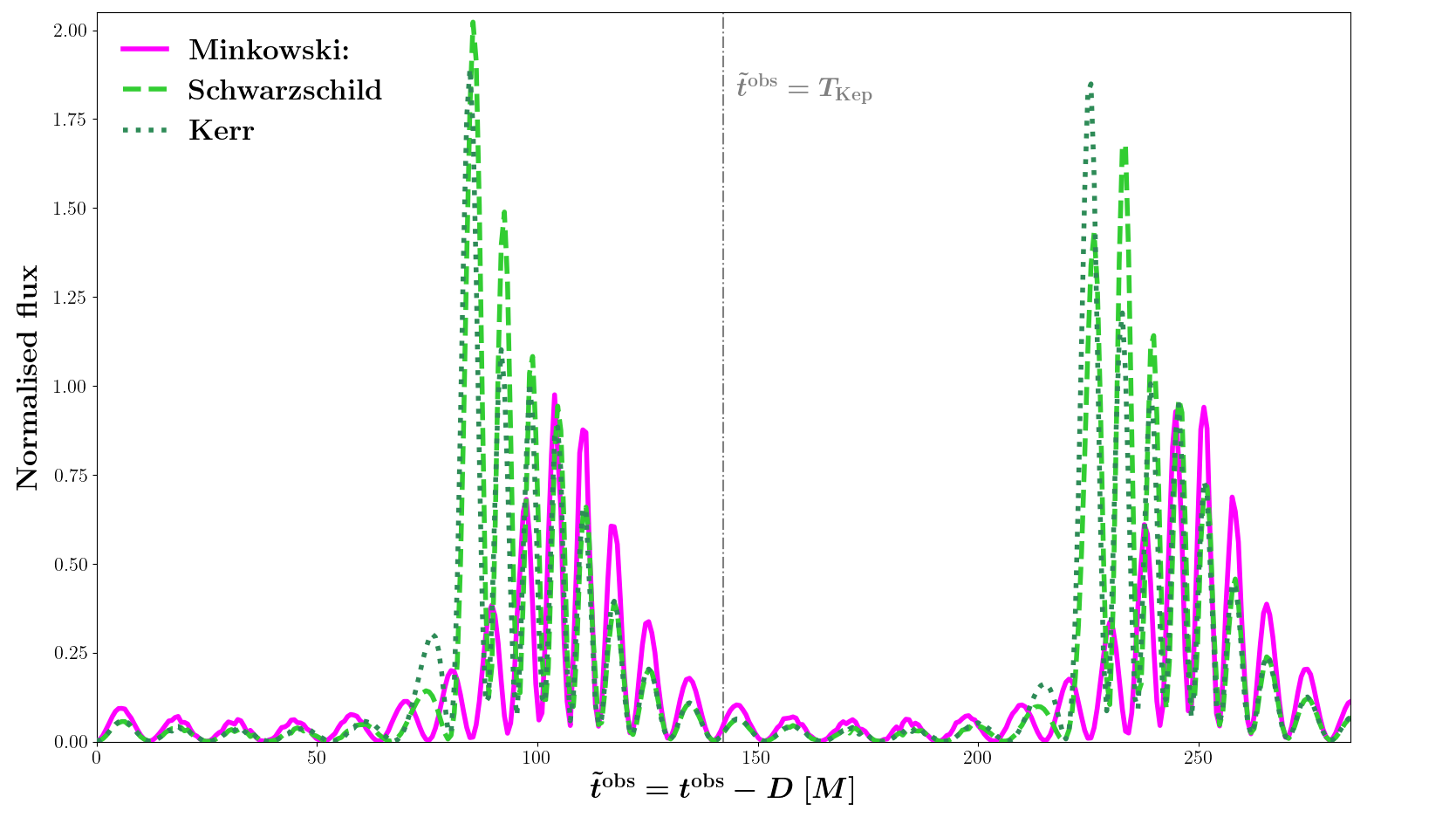}
		\caption{Normalised light curves collected at $i=100^\circ$ over two orbital periods in Minkowski (solid magenta), Schwarzschild (dashed light green), and Kerr ($a=0.94$, dotted dark green) space-times.}
		\label{fig:i100}
	\end{figure}

	\noindent As anticipated by the remarks in Appendix \hyperref[app:hit]{\ref{app:hit}}, higher inclinations exacerbate asymmetries, particularly in curved space-times. For example, the light curves for $i=100^\circ$ represented in Figure \hyperref[fig:i100]{\ref{fig:i100}} exhibit a more pronounced accordion-like behaviour due to light-travel-time delays, as discussed in Section \hyperref[sec:times]{\ref{sec:times}}, than those for $i=160^\circ$ (see left panel of Figure \hyperref[fig:starting_point]{\ref{fig:starting_point}}), as well as additional skewness in Schwarzschild and Kerr space-times, mainly driven by the areal evolution analysed in Section \hyperref[sec:area]{\ref{sec:area}}. 
	
\begin{figure}[ht!]
		\centering
		\includegraphics[width=8.5cm]{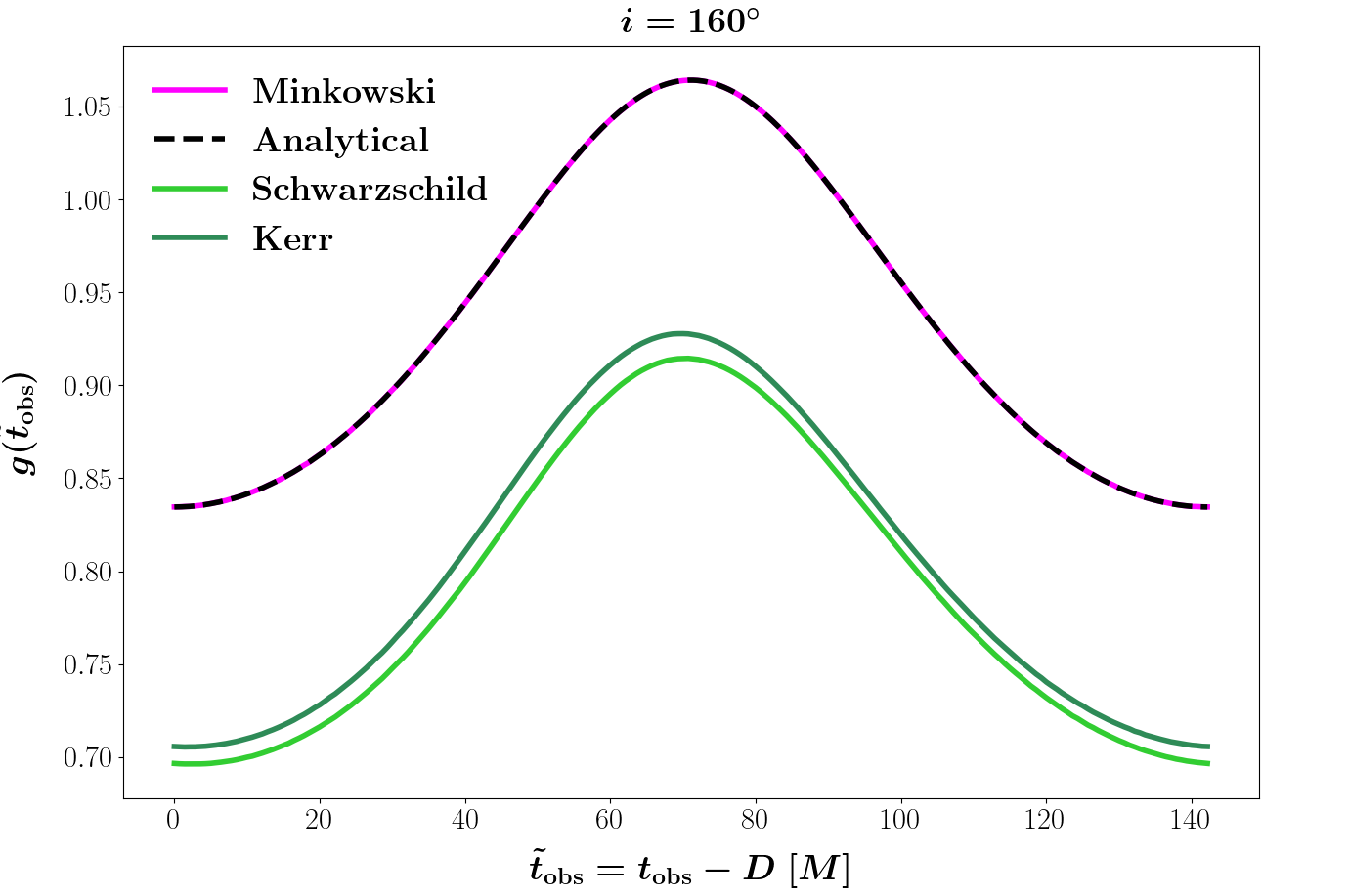}
		\caption{Redshift profile of the hot spot's centre in Minkowski (magenta), Schwarzschild (light green), and Kerr ($a=0.94$, dark green) space-times at $i=160^\circ$.  Within flat space-time, the dashed black curve represents the analytical prediction \eqref{eq:redshiftMinko}.}
		\label{fig:g160}
	\end{figure}

	Similarly, the redshift profiles for $i=160^\circ$ shown in Figure \hyperref[fig:g160]{\ref{fig:g160}} display less significant variations and retain a more symmetric shape compared with those for $i=100^\circ$ (see Figure \hyperref[fig:redshift]{\ref{fig:redshift}}).
	
	\end{appendix}
	\label{LastPage}
	
\end{document}